\pdfoutput=1
\documentclass[useAMS,usenatbib]{mn2e}
\usepackage{natbib}
\usepackage{amssymb}
\usepackage{graphicx}

\title[The non-linear post-Friedmann vector potential]{The fully non-linear post-Friedmann frame-dragging vector potential:\\
Magnitude and time evolution from N-body simulations}
\author[D. B. Thomas et al. 2014] {Daniel B.~Thomas$^{1,2}$\thanks{E-mail: daniel.b.thomas@port.ac.uk}, Marco Bruni$^1$, David Wands$^1$  \\ \\ 
  1 - Institute of Cosmology and Gravitation, University of Portsmouth, Dennis Sciama Building, Burnaby Road,\\
  \quad \, Portsmouth, PO1 3FX, UK\\
  2 - Department of Physics, University of Cyprus, Aglantzia, Nicosia, 2109}

\date{Accepted ;
  Received ; in original form }
\pagerange{\pageref{firstpage}--\pageref{lastpage}} \pubyear{2014}

\begin{document}

\maketitle

\label{firstpage}

\begin{abstract}
Newtonian simulations are routinely used to examine the matter dynamics on non-linear scales. However, even on these scales,
Newtonian gravity is not a complete description of gravitational effects. A post-Friedmann approach shows that the leading
order correction to Newtonian theory is a vector potential in the metric. This vector potential can be
calculated from N-body simulations, requiring a method for extracting the velocity field. Here, we present the full details
of our calculation of the post-Friedmann vector potential, using the Delauney Tesselation Field Estimator (DTFE) code. We
include a detailed examination of the robustness of our numerical result, including the effects of box size and mass
resolution on the extracted fields. We present the power spectrum of the vector potential and find that the power spectrum
of the vector potential is $\sim 10^5$ times smaller than the power spectrum of the fully non-linear scalar gravitational
potential at redshift zero. Comparing our numerical results to perturbative estimates, we find that the fully non-linear
result can be more than an order of magnitude larger than the perturbative estimate on small scales. We extend the analysis
of the vector potential to multiple redshifts, showing that this ratio persists over a range of scales and redshifts. We
also comment on the implications of our results for the validity and interpretation of Newtonian simulations.
\end{abstract}


On the largest scales in cosmology, theoretical calculations can be carried out using standard cosmological perturbation theory.
These calculations fully encompass General Relativity (GR) but are limited to scales where the perturbations, in particular the
density perturbation, are small. On smaller scales, where the focus is on non-linear structure formation, Newtonian N-body
simulations are used. These simulations do not require that the density contrast is small, but they suffer from the limitations
of being Newtonian rather than GR simulations. There is an entire field in cosmology dedicated to developing, running and
analysing these Newtonian N-body simulations. There has been sporadic interest in understanding the use of Newtonian theory in
cosmology \citep{tomitaflat,asada,takfut97,mataterra,carbone,hwangpn,hwangnonlin,postf,1207.2035,1211.0011,1312.3638}, as well as
examining the relativistic interpretation of the simulations \citep{chis11,green12,Bruni:2013mua,1308.6524}. These studies have
predominantly focussed on whether the dynamics of density contrast and scalar potential accurately match those of GR.

In this paper, we are mostly interested in another important limitation of Newtonian simulations: Even if the matter dynamics are
being computed correctly, there are cosmological quantities of interest on non-linear scales that have no counterpart in
Newtonian theory. Examples of these quantities include the difference between the two scalar potentials, gravitational waves and
the vector potential in the metric, all of which must exist on non-linear scales in a GR universe.

These extra quantities would naively be expected to be small if the Newtonian simulations are a good approximation to a GR
universe. However, explicitly calculating these quantities has several advantages: To start with, it would be good to have a
quantitative check of whether these quantities are small, and indeed how small they are. In particular, as we enter the era of
precision cosmology, we need to check that these quantities will not affect the observables at the percent-level. Furthermore,
checking that these quantities are negligible provides a quantitative check on the Newtonian approximation in a $\Lambda$CDM
cosmology.

We will be working with the post-Friedmann formalism \citep{postf,thesis}. This generalises to cosmology the weak-field
(post-Minkowski) approximation, with a post-Newtonian style expansion \citep{chand,weinbook,poisandwillbook} in inverse powers of
the speed of light $c$ of the perturbative quantities. These expansions need to be performed differently in cosmology compared to in the
Solar System due to the different situations and aims in the two cases. For example, the time-time and space-space components of
the metric need to be treated at the same order in cosmology in order for the resulting equations to be a consistent solution of
the Einstein equations.

The post-Friedmann formalism, when linearised, correctly reproduces conventional linear perturbation theory and can thus describe
structure formation on the largest scales. More importantly, the leading order equations in the $1/c$ expansion can be examined
and are expected to yield the non-linear Newtonian equations. Note that in this ``Newtonian'' regime, the density contrast has not
been assumed to be small. The equations in this regime will be shown in section \ref{sec_postf}, essentially comprising the
Newtonian equations, as expected, plus an additional equation. This additional equation shows how the vector potential in the
metric, the lowest order beyond-Newtonian quantity, is generated by the matter dynamics. This vector potential is the
beyond-Newtonian quantity that we will examine in this paper, it is the cosmological manifestation of the ubiquitous relativistic
effect of frame dragging. This effect has been measured in the Solar System by Gravity Probe B \citep{gpb}.

In this paper, we present a calculation of this vector potential based on extracting the density and velocity fields from N-body
simulations. We expand on the results of \cite{Bruni:2013mua}, which was the first calculation of an intrinsically relativistic
quantity on fully non-linear scales from large scale cosmological matter fields, rather than from individual astrophysical
occurences. The main focus in this paper is to present the method used to extract this vector potential from N-body simulations.
In particular, we examine the robustness of the numerical extraction of the vector potential and present the tests we carried out
to examine the numerical effects of simulation parameters on the extraction, which were not presented in \cite{Bruni:2013mua}.

The main physical results of this paper are figures \ref{fig_gravz0} and \ref{fig_gravredshifts}, showing the power spectrum of the
vector potential at redshift zero and its evolution with time respectively. Additionally, we have presented the ratio of the vector
potential power spectrum to that of the scalar potential in figures \ref{fig_potratioz0_err} and \ref{fig_ratioredshifts}. The
results on the magnitude and evolution of the power spectrum of the vector potential in this paper were used in \cite{1403.4947} to
examine the possible weak-lensing consequences of the vector potential.

This paper is laid out as follows. In section \ref{sec_postf}, we present the pertinent details of the post-Friedmann formalism and
show the equation governing the vector potential. We will also present our definitions and notation regarding vector power spectra.
In section \ref{sec_sims}, we explain how the relevant fields were extracted from N-body simulations and examine the robustness of this
extraction. In section \ref{sec_results}, we show the power spectrum of the vector gravitional potential and its time evolution, as well
as comparing it to the closest analytical results in the literature. We conclude in section \ref{sec_conc}. Appendix \ref{app_vec}
contains some details about vector power spectra and in Appendix \ref{app_robust} we show results from some of the numerical tests that
were carried out. Additional plots are available as online supplementary material, divided amongst three files:
Resolution\_and\_BoxSize\_Dependence.pdf (hereafter RB), GridSize\_and\_Binning\_Dependence.pdf (hereafter GB) and
ConsistencyChecks.pdf (hereafter CC).

\section{Post-Friedmann Formalism}
\label{sec_postf}
The post-Friedmann approach is developed in \cite{postf} and \cite{thesis}, see there for the full details. This approach considers a dust
(pressure-less matter) cosmology with a cosmological constant. The perturbed FLRW metric, in Poisson gauge, is expanded up to order
$c^{-5}$, keeping the $g_{00}$ and $g_{ij}$ scalar potentials at the same order:
\begin{eqnarray}
 g_{00}&=&-\left[1-\frac{2U_N}{c^2}+\frac{1}{c^4}\left(2U^2_N-4U_P \right)\right]\nonumber\\
g_{0i}&=&-\frac{aB^N_i}{c^3}-\frac{aB^P_i}{c^5}\\
g_{ij}&=&a^2\left[\left(1+\frac{2V_N}{c^2}+\frac{1}{c^4}\left(2V^2_N+4V_P \right) \right)\delta_{ij} +\frac{h_{ij}}{c^4}\right]\nonumber
\end{eqnarray}
The $g_{00}$ and $g_{ij}$ scalar potentials have been split into the Newtonian ($U_N$, $V_N$) and post-Friedmann ($U_P$, $V_P$) components.
Similarly, the vector potential has been split up into $B^N_i$ and $B^P_i$. Since this metric is in the Poisson gauge, the three-vectors
$B^N_i$ and $B^P_i$ are divergenceless, $B^N_{i,i}=0$ and $B^P_{i,i}=0$. In addition, $h_{ij}$ is transverse and tracefree,
$h^i_i=h_{ij}^{,i}=0$. Note that at this order, $h_{ij}$ is not dynamical, so it does not represent gravitational waves. From a
post-Friedmann viewpoint, there are two different levels of perturbations in the theory, corresponding to terms of order $c^{-2}$ and
$c^{-3}$, or of order $c^{-4}$ and $c^{-5}$ respectively. Defining ``resummed" variables, such as $\Phi=2U_N+c^{-2}\left(2U^2_N-4U_P \right)$,
then calculating the Einstein equations and linearising them, reproduces linear GR perturbation theory in Poisson gauge. Thus, this approach
is capable of describing structure formation on the largest scales.

For smaller scales, in a dust cosmology, we are interested in the weak field, slow motion, sub-horizon, quasi-static and negligible pressure
regime. This is simply derived by retaining only the leading order terms in the $c^{-1}$ expansion and upon doing so we recover Newtonian
cosmology, albeit with a couple of subtleties. The first is that the space-time metric is a well-defined approximate solution of the
Einstein equations. The second is that we have an additional equation, which is a constraint equation for the vector gravitational potential
$B^N_i$. The full system of equations obtained from the Einstein and hydrodynamic equations \citep{postf}, given the evolution of the
background $a(t)$, is as follows.
\begin{eqnarray}
&&\frac{d\delta}{dt}+\frac{v^i_{,i}}{a}\left( 1+\delta \right)=0\label{eq_newt1}\\
&&\frac{dv_i}{dt}+\frac{\dot{a}}{a}v_i=\frac{1}{a}U_{N,i}\label{eq_newt2}\\
&&\frac{1}{c^2a^2}\nabla^2V_N=-\frac{4\pi G}{c^2}\rho_b \delta\label{eq_newt3}\\
&&\frac{2}{c^2a^2}\nabla^2\left(V_N-U_N \right)=0\\
&&\frac{1}{c^3}\hspace{-0.1cm}\left[\frac{2\dot{a}}{a^2}U_{N,i}+\frac{2}{a}\dot{V}_{N,i}-\frac{1}{2a^2}\nabla^2B^N_i\right]\hspace{-0.1cm}=
\hspace{-0.1cm}\frac{8\pi G\rho_b}{c^3}\left(1+\delta \right)v_i\label{eq_vector1}
\end{eqnarray}
As expected, we have the Newtonian continuity and Euler equations from the hydrodynamic equations as well as Poissons equation from the Einstein
equations. Note that the time derivative here is the convective derivative, $dA/dt=\partial A/\partial t+v^i A_{,i}/a$, for any quantity $A$.
The Einstein equations yield two additional equations: The first is an equation forcing the scalar potentials $V_N$ and $U_N$ to be equal,
consistent with there being only one scalar potential in Newtonian theory. Note that some approaches consider the potentials to be a priori equal
at leading order whereas here we assumed the full generality of GR and the equality of the potentials arose naturally on taking the Newtonian regime.
The second additional equation relates the leading order vector gravitiational potential, $B^N_i$, to the momentum of the matter. Thus, even in the
regime where the matter dynamics are correctly described by Newtonian theory, the frame-dragging potential $B^N_i$ should not be set to zero; this
would correspond to putting an extra constraint on the Newtonian dynamics. We note that there is a similar equation in several other formalisms in
the literature \citep{takfut97,hwangnonlin,green12}.
We can see from equation (\ref{eq_vector1}) that the potential $B^N_i$ is sourced by the vector part of the energy current $\rho \vec{v}$. This is
made apparent by taking the curl of this equation, which gives
\begin{equation}
\label{eqn_curl}
 \nabla \times \nabla^2 \vec{B}^N=-\left(16\pi G \rho_b a^2\right)\nabla\times\left[(1+\delta)\vec{v} \right]\rm{,}
\end{equation}
where the source term on the right hand side splits up into three terms: the vorticity $\nabla \times \vec{v}$ and then two further terms,
\begin{equation}
\label{eqn_threecomp}
\nabla\times\left[(1+\delta)\vec{v} \right]=\nabla \times \vec{v}+\delta\nabla\times \vec{v}+\nabla\delta\times\vec{v}\rm{.}
\end{equation}

It is equation (\ref{eqn_curl}) that will be used for the rest of the paper. Since the matter dynamics are not affected at this order, i.e. they are described
by the standard Newtonian equations (\ref{eq_newt1}-\ref{eq_newt3}), the density and velocity fields sourcing the vector potential are Newtonian
and can be extracted from N-body simulations. 
Using the definitions of vector power spectra in Appendix \ref{app_vec}, the power spectrum of the vector potential is given by
\begin{equation}
P_{{\vec B}^N}(k)=\left(\frac{16\pi G \rho_b a^2 }{k^2}\right)^2\frac{1}{k^2}P_{\delta v}(k) \rm{,}
\end{equation}
with
\begin{eqnarray}
&&P_{\delta v}=P_{\nabla \times {\vec v}}(k)+P_{\delta \nabla \times {\vec v}}(k)+P_{(\nabla \delta) \times {\vec v}}(k) \\
&&+P_{\left(\nabla \delta \times \vec{v}\right)\left(\nabla \times \vec{v}\right)}(k) +P_{\left(\nabla \delta \times \vec{v}\right)
\left( \delta\nabla \times \vec{v}\right)}(k)
+P_{\left(\delta\nabla \times \vec{v}\right)\left(\nabla \times \vec{v}\right)}(k)\nonumber\rm{.}
\end{eqnarray}
Unless stated otherwise, all plots of the gravitational potentials show the dimensionless power spectrum $\Delta(k)$, see Appendix \ref{app_vec} for conventions.

\section{Simulations}
\label{sec_sims}
Our simulations have all been run using the publicly available N-body code Gadget2 \citep{gadget2}. Many simulations have been run in order to
quantify the effects of box size and mass resolution on the quantities that we are extracting, see table \ref{table_sims} for a full list of the
simulations. All of the simulations were run with dark matter particles only, as the equation for the vector potential is derived for a
pressureless matter and cosmological constant cosmology. To allow comparison to previous studies of vorticity \citep{pueb}, the simulations were
run with the cosmological parameters $\Omega_m=0.27$, $\Omega_{\Lambda}=0.73$, $\Omega_b=0.046$, $h=0.72$, $\tau=0.088$, $\sigma_8=0.9$ and $n_s=1$.
All of the simulations started at redshift 50 and had their initial conditions created using 2LPTic \citep{2lptb}. Our final result for the vector
potential is taken from the three $160h^{-1}$Mpc simulations with $1024^3$ particles, these will be referred to as the high-resolution (HR) simulations.

\begin{table}
 \caption{Parameters for the simulations.}
\centering \begin{tabular}{|c| c| c| c|c|}
\hline
Box size& Particle & Mass resolution& Number of & Softening\\
($h^{-1}$Mpc) & number & ($10^{8}$ M$_{\odot}$) & Realisations& ($h^{-1}$kpc)\\
\hline
 80 &  512$^3$ & 3.97 & 8  & 6.25\\
 80 &  512$^3$ & 3.97 & 1  & 4.0 \\
140 &  768$^3$ & 6.31 & 8 & 6.25\\
140 &  560$^3$ & 16.3 & 8  & 6.25\\
160 & 1024$^3$ & 3.97 & 3  & 6.25\\
160 &  880$^3$ & 6.26 & 3 & 6.25\\
160 &  640$^3$ & 16.3 & 8  & 6.25\\
160 &  640$^3$ & 16.3 & 1  & 5.0 \\
160 &  320$^3$ &  130 & 8  & 15.0\\
200 & 1024$^3$ & 7.76 & 2 & 6.25\\
240 &  960$^3$ & 16.3 & 3 & 6.25\\
240 &  480$^3$ &  130 & 8  & 15.0\\
320 &  640$^3$ &  130 & 8  & 15.0\\
\hline
\end{tabular}
\label{table_sims}
\end{table}

\begin{table}
 \caption{Redshifts used to probe time evolution of quantities.}
\centering \begin{tabular}{|c| c| c|}
\hline
Scale Factor & Redshift & Colour on time evolution plots \\
\hline
0.33 &  2.0  & black \\
 0.4 &  1.5  & red \\
 0.5 &  1.0  & magenta \\
 0.6 &  0.67 & yellow \\
 0.7 &  0.43 & green \\
 0.8 &  0.25 & cyan \\
 0.9 &  0.11 & blue \\
 1.0 &  0.0  &  brown \\
\hline
\end{tabular}
\label{table_redshifts}
\end{table}

\subsection{Tesselation}
To extract the necessary fields from the simulations, the Delauney Tesselation Field Estimator (DTFE) code was used \citep{dtfecode}. Standard
methods of extracting fields from N-body simulations, such as Cloud-In-Cells (CIC) \citep{cic} work well for the density field, as the particles,
by definition, sample the density field well. However, these methods have several shortcomings when applied to the extraction of velocity fields:
One is that the field is only sampled where there are particles, so in a low density region the velocity field is artificially set to zero. In
addition, the extracted field will be a mass-weighted, rather than volume-weighted field. A consequence of these shortcomings is that, as the
grid size is increased, the velocity field will not converge. In fact, it will become zero in an increasing proportion of the grid cells as the
grid size increases. Several authors have looked at using the Delauney tesselation \citep{dtfe1,dtfe2,dtfe3} for astrophysical applications
including the examination of velocity fields. See also \cite{pueb} for comparisons of extracting velocity fields with tesselations rather than
more standard methods. The DTFE code constructs the Delauney tesselation of the set of particles, consisting of tetrahedra whose nodes are located
at the particles positions. The tetrahedra are constructed such that the circumsphere of each tetrahedron does not contain any of the particles
except for the particles located at the nodes of the tetrahedron in question. This makes the tesselation unique. The particles' velocities are then
linearly interpolated across each tetrahedron, yielding a value for the smoothed velocity field and its gradients at every point in the simulation
volume. A regular N$^3_{\rm{grid}}$ grid is laid down and the code samples $N_{\rm samples}$ points at random in each grid cell and averages the
field over these points, giving a value for the smoothed field in each grid cell.
Once the fields are obtained on the regular grid, the power spectra are calculated using the standard process of averaging the modulus-squared of
the Fourier coefficients over a given range of $k$. For the analyses here, we used $N_{\rm grid}/4$ bins, although varying this value does not
affect the results (see Appendix \ref{app_bins}).

\subsection{Convergence and Tests}
It is important to ensure that our numerical result for the vector potential is robust and independent of the simulation parameters. In this subsection
we will present the results of our examination into the effects of different simulation parameters on the extracted vector power spectrum. Since the
velocity and density fields both contribute to the source for the vector potential, we will examine the density, vorticity and velocity divergence spectra
too: We will examine their behaviour individually, compare them to other studies and methods of extraction and also consider the consistency of the
extracted fields through the relations
\begin{eqnarray}
k^2P_{\delta}(k)&=&P_{\nabla \delta}(k)\nonumber\\
k^2 P_{\vec{v}}(k)&=&P_{\nabla \cdot \vec{v}}(k)+P_{\nabla \times \vec{v}}(k) \rm{.}
\end{eqnarray}

The box size and mass resolution of the simulation are the two main parameters whose effect on the extracted fields needs to be examined. In addition,
we have examined the effect of varying the grid size and $N_{\rm samples}$, which are both internal DTFE parameters. The parameters of the different
simulations used are in table \ref{table_sims}. We chose the softening lengths of the N-body simulations to be consistent with \cite{pueb} in order
to recreate their study of the velocity divergence and vorticity, however varying the softening length did not influence the results, see Appendix \ref{app_soft}.

Although we did run some simulations with a box size below 140$h^{-1}$Mpc, we have not included these in the analysis here as smaller box sizes have
systematically less power. See Appendix \ref{app_smallboxes} for the results from these simulations and how they compare to the larger box sizes. For
further results regarding the effects of a small simulation box on cosmological quantities, see \citep{0601320,0410373,9408029}.

\subsubsection{A note on error bars}
Since we have only three realisations of our HR simulations, we cannot compute meaningful error bars. Thus, we have not included any error bars on the
majority of our plots. Instead, in figures \ref{fig_deltavcurl}, \ref{fig_deltavcic}, \ref{fig_potratioz0_err} and \ref{fig_deltavratioz0_err}, we have
plotted the results from the three individual realisations, in order to illustrate by how much the results vary. Unless stated otherwise, the results shown in
the other plots show the average over the realisations. We explicitly examine the variation amongst realisations in Appendix \ref{app_real}
for several quantities, notably the vorticity and vector potential. In particular, we note there that when considering the vector potential, cosmic variance
on the largest scales affects smaller scales, as explained by a perturbative analysis \citep{0709.1619,0812.1349}. See \ref{app_real} for more discussion of
this. We also note there that the variation of the vorticity amongst realisations seems to be larger than the variation of the density, although there seems to be
no discussion of this in the literature.

\subsubsection{Mass Resolution}
We have examined the dependence of the density, velocity divergence, vorticity and vector potential on the mass resolution of the simulations. For the
density and velocity divergence there is evidence for a mild dependence on mass resolution for both of these fields on smaller scales. This is likely to be
due to the DTFE window function, which cannot be compensated for, rather than a mass-resolution dependence of the field itself. There is no evidence of any
mass-resolution dependence of these fields on larger scales. The variation of the density and velocity divergence with mass resolution can be seen in figures
1 and 2 of file RB. The effect of the small-scale mass-resolution dependence is negligible for our HR simulations, as seen when comparing to alternative
methods of calculating the density power spectrum.

The dependence of the vorticity power spectrum with mass resolution is shown in figure \ref{fig_vortresn}. The power spectrum shows spurious additional
power when the mass resolution is insufficient. However, once the resolution is sufficient, of order $10^{9}M_{\odot}$, there is no evidence for any
systematic dependence on mass resolution. This dependence on mass resolution, followed by convergence around $\sim10^{9}M_{\odot}$, matches previous
findings, notably those of \cite{pueb}.

In figure \ref{fig_gravresn}, we show the dependence of the vector potential on mass resolution. There is a clear dependence of the vector potential
on mass resolution, similar to that seen for the vorticity. However, there are several differences. In particular, the mass-resolution dependence seems
to be less important for smaller scales, where there is a greater dependence on box size (see later). In addition, the dependence on mass-resolution is
still apparent around $10^{9}M_{\odot}$. However, once there mass resolution has improved to around $6\times10^{8}M_{\odot}$, there is no evidence of a
mass resolution dependence of the vector potential.

To show this further, figure \ref{fig_gravratiovsreal} shows the higher resolution simulations in more detail, complete with the individual realisations
of the HR simulations. The y-axis here is $k^2 P_{\vec{B}}(k)$ in order to show the variance more clearly over the range of scales being considered.
The cyan line shows the simulation with the worst resolution ($16.3\times10^{8}M_{\odot}$) of those in this plot and indeed this simulation shows a
systematic deviation on the largest scales. The better resolution simulations show better convergence, with the 140h$^{-1}$Mpc simulation with 768$^3$
particles being consistent with the HR simulations for essentially the entire range under consideration. This convergence is examined further in Appendix
\ref{app_real}.

\subsubsection{Box Size}
We have also considered the effect of varying the box size on the extracted power spectra. As expected, there is no evidence for any
systematic dependence of the density, vorticity and velocity divergence power spectra on the box size of the simulations. This can be
seen in figures 3, 4 and 5 of file RB. Note that, for sufficiently small boxes, a systematic deviation can arise, see
Appendix \ref{app_smallboxes}.

Figure \ref{fig_gravbox} shows the box size dependence of the vector potential. As mentioned above, the vector potential does show some dependence on box size.
The vector potential shows signs of a dependence on the box size on scales below 1$h^{-1}$Mpc, however this is difficult to entangle from the effects of mass
resolution and the window function. For box sizes below 200$h^{-1}$Mpc, there is no systematic dependence of the vector potential power spectrum with box size.\\
In Appendix \ref{app_real}, we examine the variation between realisations for the vector potential, and relate it to the behaviour that might be expected from
perturbative arguments. In particular, figure \ref{fig_gravratiovsreal} shows how the variation between realisations is larger than the effects of box size and
mass resolution for simulations with box sizes below 200$h^{-1}$Mpc and mass resolution of at least $6\times10^{8}M_{\odot}$. Thus, we expect numerical effects
from the simulation parameters to be a sub-dominant source of error as long as the parameters are within this range.

\subsubsection{Consistency Checks}\label{sec_consist}
There are a few consistency checks that can be performed on the different fields that we are interested in. The quantities that are used for the vector potential
include the density field and its gradients as well as the velocity field and its gradients. There are two relations between these fields and their derivatives,
\begin{eqnarray}
\label{eq_consist}
k^2P_{\delta}(k)&=&P_{\nabla \delta}(k)\\
k^2 P_{\vec{v}}(k)&=&P_{\nabla \cdot \vec{v}}(k)+P_{\nabla \times \vec{v}}(k)\rm{.}
\end{eqnarray}
We have extracted the quantities on the left and right sides of these relations from our HR simulations and compared them, see figure 1 in file CC for the ratio
$P_{\nabla \delta}(k)/k^2P_{\delta}(k)$ and figure 2 in file CC for the ratio
$k^2 P_{\vec{v}}(k)/(P_{\nabla \cdot \vec{v}}(k)+P_{\nabla \times \vec{v}}(k))$.
In both cases, two curves are plotted, corresponding to two different methods of calculating the ratio. The blue line shows the ratio exactly as suggested
above, with the factor of $k$ in equation \ref{eq_consist} taken to be the value defining the centre of the bin. For the red curve, the exact k-value for each
mode is used when computing the sum in each bin. For small bins, or fields where the values vary slowly as a function of $k$, these two should agree and
indeed they do for smaller scales where our (logarithmic) bins are smaller. There is a difference between the methods for the largest scales in our simulations,
this will be discussed below for each test.

For the density field, the two methods for calculating the ratio do give different answers. However, for both methods, the deviation is within 2\% for every
bin except the first. Thus, this consistency check for the density field is well satisfied for all scales $k\geq0.2 h$Mpc$^{-1}$.

The consistency check for the velocity field is less well satisfied: there is a sharp divergence in the power spectra on the smallest scales, such that the check
is not satisfied within $10\%$ at $k\approx8h^{-1}$Mpc. This shows the effect of the DTFE window function on the extracted fields. We will not consider the
extracted vector potential for k larger than $k\approx8h^{-1}$Mpc when presenting our results. Furthermore, the two methods show very different behaviour:
the method using the average $k$-value for each bin causes the consistency test to fail on large scales. However, with the more exact method, the consistency
check is very well satisfied on all of the largest scales. This suggests that the dominant contribution to the bins on the largest scales comes from the
low-$k$ end of each bin, hence the overestimation of $k^2P_{\vec{v}}(k)$ when the average $k$-value for each bin is used. The strong effect here is partly
caused by the relatively steep slope of the velocity power spectrum. We note that this effect would also come into play when calculating the dimensionless
velocity power spectrum for binned data. Nonetheless, the good agreement of the consistency check when using the second method is strong evidence that the
derivatives of the velocty field are being calculated correctly.

\begin{figure}
\begin{center}
\includegraphics[width=2.7in,angle=270]{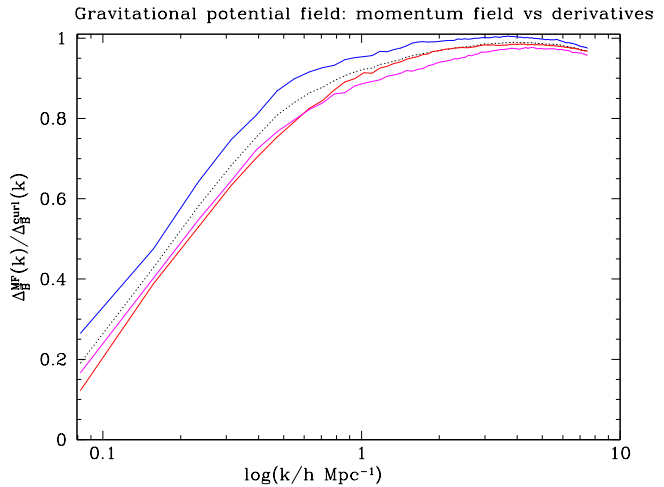}
\end{center}
\caption{The ratio of the vector potential power spectra computed using the vector part of the momentum field and the curl of the momentum field. The blue,
magenta and red curves show the ratio for the three realisations of the HR simulations, and the black (dashed) curve shows the average over these three. There is
reasonable agreement between the two power spectra for the smaller scales, however the two methods diverge for the largest scales and there is a difference
of a factor of 5 at the largest scales. For most of the range of $k$ under consideration ($k\geq0.2h$Mpc$^{-1}$), the two vector power spectra agree to within
a factor of 2.}
\label{fig_deltavcurl}
\end{figure}

A further check that we can perform is to extract the complete momentum field, $\vec{p}=(1+\delta)\vec{v}$, and decompose it into its vector and scalar parts
directly rather than dealing with derivatives. The power spectrum of the vector potential can then be calculated from the vector part of the momentum field,
$\vec{p}^v$, using
\begin{equation}
P_{{\vec B}^N}(k)=\left(\frac{16\pi G \rho_b a^2 }{k^2}\right)^2 P_{\vec{p}^v}(k) \rm{.}
\end{equation}
In figure \ref{fig_deltavcurl} we show the ratio of the vector power spectrum calculated using the two methods, with the different lines corresponding to
different individual realisations. The vector potentials calculated from the two methods are broadly consistent, within 20\% for most of the range under
consideration, and agreeing to within a factor of 2 for $k\geq0.2h$Mpc$^{-1}$. We are unsure what the causes of the difference between the two methods are.
In particular, we checked for whether there is an effect coming from the use of $k$ averaged over the bin, as in the velocity field consistency check, however
this effect is negligible for the gravitomagnetic potential\footnote{As an aside, we note that we also calculated the momentum field by extracting the velocity
field and density field separately at each grid point, before multiplying them together. The power spectrum calculated from this field agrees well with that
calculated by extracting the momentum field as a single field. The same agreement is not obtained when extracting the field $\delta^2$ and comparing to squaring
the density field, when using either the DTFE code or a CiC method.}. The difference between the methods is larger than the variation amongst realisations for
either method.

\begin{figure}
\begin{center}
\includegraphics[width=2.7in,angle=270]{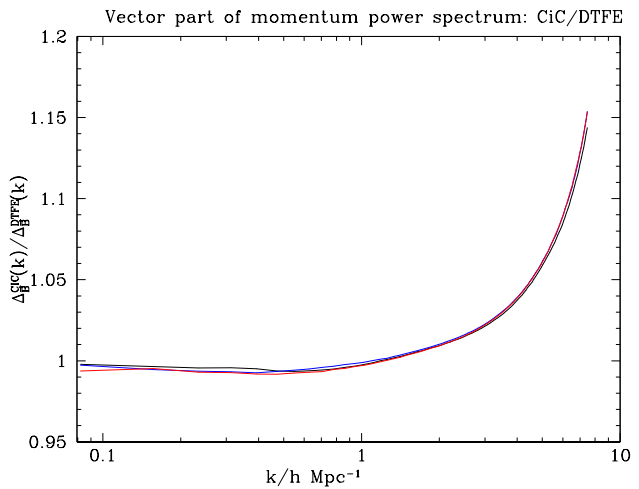}
\end{center}
\caption{The ratio of the vector potential power spectra computed using the vector part of the momentum field calculated using the Cloud-in-Cells method and
the DTFE method. The blue, magenta and red curves show the ratio for the three realisations of the HR simulations, and the black curve shows the average over
these three. The two methods agree very well on larger scales, but diverge for the smallest scales.}
\label{fig_deltavcic}
\end{figure}

We can also extract the momentum field directly using a standard cloud-in-cells (CiC) approach \citep{cic}, and compare this to the momentum field extracted
using the DTFE code. The ratio between these fields is shown in figure \ref{fig_deltavcic}. There is good agreement between the two methods of computing the
momentum power spectrum on larger scales, but with a divergence between the two methods on smaller scales. It is unclear which method would be expected to be
more accurate on these smaller scales: the DTFE method suffers from having a window function that cannot be deconvolved, however the CiC method will have cells
with a zero momentum field, due to the lack of nearby particles, for a sufficiently large grid. In fact, the CiC method does not converge as the grid size is
increased. We used a $512^3$ grid for the CiC code, although we checked that changing this to $256$ or $1024$ does not significantly affect the results. Unlike
the DTFE method, derivatives cannot be directly extracted with the CiC method, so the consistency checks performed earlier for the DTFE method cannot be applied
to the CiC method. This also means that the first method of extracting the vector potential, using the curl of the momentum field, cannot be carried out with the
CiC method.

We present the vector power spectrum from both the momentum field and the curl method in the results section. We note that the level of agreement between figures
\ref{fig_deltavcurl} and \ref{fig_deltavcic} suggests that our vector potential power spectrum is robust and correct to within a factor of 2. It is unclear to us which method should
be trusted more; whilst the momentum field method is simpler, the derivative method allows us to examine the different components, notably the vorticity, and
check that it behaves as expected. The differences between the two methods do not affect the observability of the vector potential, see \cite{Bruni:2013mua} and
\cite{1403.4947}.

\subsubsection{Comparison to previous findings}
There are several works in the literature to which we can compare our findings on the velocity field and its components. As mentioned above, the vorticity
and velocity divergence power spectra were extracted from N-body simulations in \cite{pueb} using an alternative implementation of the Delauney tesselation.
They found a strong dependence on resolution of the extracted vorticity power spectrum and an approximate scaling of the vorticity power spectrum with the
seventh power of the linear growth factor.

The vorticity and velocity divergence power spectra in \cite{pueb} are consistent with the spectra extracted for this paper and we found the same resolution
dependence of the vorticity power spectrum (see above). However, as detailed in Appendix \ref{app_linear}, we do not find the same scaling of the vorticity
spectrum with the seventh power of the linear growth factor ($D_+$): Although this scaling seems to hold at low redshift, it no longer holds at redshift one and beyond. At
these earlier times, the power spectrum is smaller than expected from the growth factor to the seven scaling, so the vorticity power spectrum must have grown
by less at redshift two than expected. 

Two recent publications \citep{1308.0886,1312.1022} have examined the velocity field from the point of view of redshift space distortions. In these works,
a different method of extracting velocity fields is used, the nearest particle method. In this method, the velocity at a grid point is given by the velocity
of the nearest particle to that grid point. See those works for comments on the differences between the nearest particle and Delauney tesselation methods
of extracting the velocity power spectra. Here, we note that there appear to be pros and cons to both methods, with no clear ``better'' method. It would be
interesting to examine how close the agreement between the vector potentials extracted by the DTFE and nearest particle methods is.

Nonetheless, there are some general observations that can be compared between these works. Notably, the magnitude of the velocity and vorticity spectra is
found to be similar, considering the differences in cosmological parameters. Also, the onset of non-linearity is found to occur at lower $k$ for the velocity
divergence than for the density. In addition, \cite{1308.0886} finds a strong dependence of the curl component of the velocity field on the resolution,
similarly to both this paper and \cite{pueb}. They also find a time dependence of this component that is approximately $D^7_{+}$ up to $z=2$, although this
relationship breaks down by up to a factor of two for certain redshifts and scales. As mentioned above, whilst our simulations also find this time dependence
of the vorticity at low redshift, we find that the relationship breaks down for $z>1$. There is no examination of multiple realisations in \cite{1308.0886}
and, similarly to the comments in Appendix \ref{app_linear} regarding \cite{pueb}, the difference between our realisations is sufficient to explain the
difference between our results and those of \cite{1308.0886}.

The broad agreement between different methods, including agreement regarding resolution dependence and convergence, is promising. Details of the vorticity
field and its evolution require further study, but the vorticity is a sub-dominant contribution to the vector potential. As the simulations and snapshots
used in the papers mentioned in this section are different to ours, it is not possible to compare the methods and extracted fields any more precisely. We
note that the three works mentioned here do not have multiple realisations of their high resolution simulations, so we are unable to determine if the
variation in vorticity between realisations found by us is reproduced (see Appendix \ref{app_real}).

As this manuscript was being prepared, \cite{1404.2280} appeared on the arxiv. This paper investigates the properties of velocity divergence and vorticity
and confirms many of the findings of \cite{pueb}. In particular, they agree with our results regarding the convergence of the DTFE code for sufficient mass
resolution and our finding of a resolution dependence of the velocity divergence, which did not appear in \cite{pueb}. They use a different method to compute
the vorticity and velocity divergence power spectra, which agrees with the DTFE code for sufficient resolution. However, as with the previous papers, there
seems to be no examination of multiple realisations with the same resolution, in order to compare our findings. In addition, there is no examination of the
time dependence and thus no confirmation or rejection of the $D^7_{+}$ scaling of the vorticity spectrum at higher redshifts.

\section{Results}
\label{sec_results}
In this section we present the power spectrum of the post-Friedmann vector potential as calculated from N-body simulations. We show the power spectrum at
$z=0$ and the different components of the source, as well as the evolution of the power spectrum between $z=2$ and $z=0$. In addition, we show the ratio
between the vector and scalar power spectra, and examine the time evolution of this quantity as well. The power spectra plotted for the scalar and vector
gravitational potentials are the dimensionless power spectra. The closest analytic result to our calculation is the second order perturbative vector potential
calculated in \cite{0812.1349}. We will compare our results to theirs at redshift $z=0$, as well as comparing the time evolution.

\subsection{Results at redshift zero}\label{sec_z0}
\begin{figure}
\begin{center}
\includegraphics[width=2.7in,angle=270]{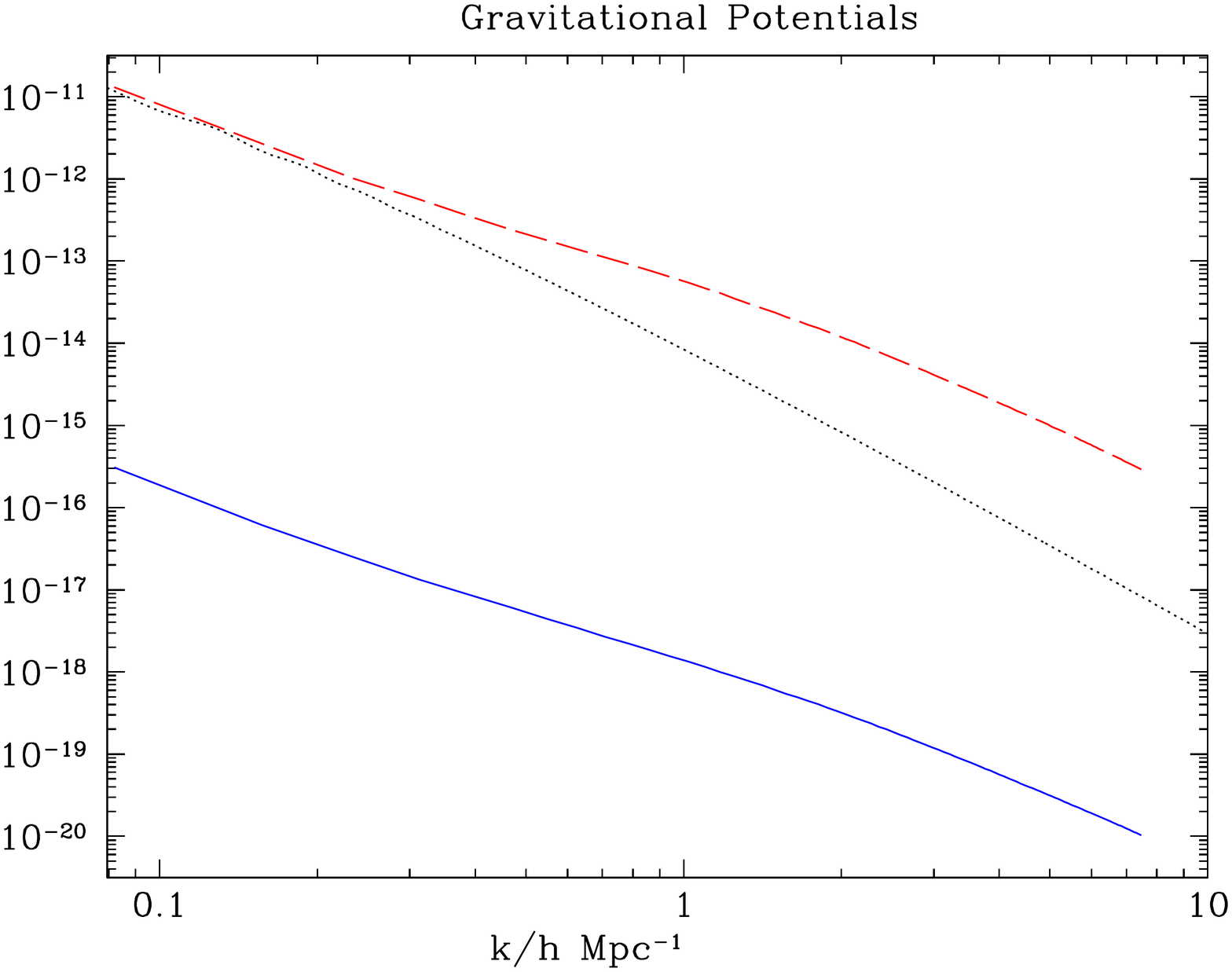}
\end{center}
\caption{The scalar (dashed red line) and vector (solid blue line) gravitational potential power spectra at redshift zero, with the vector potential
calculated using the curl method. The linear theory scalar potential is shown for comparison (dotted black line).}
\label{fig_gravz0}
\end{figure}

\begin{figure}
\begin{center}
\includegraphics[width=2.7in,angle=270]{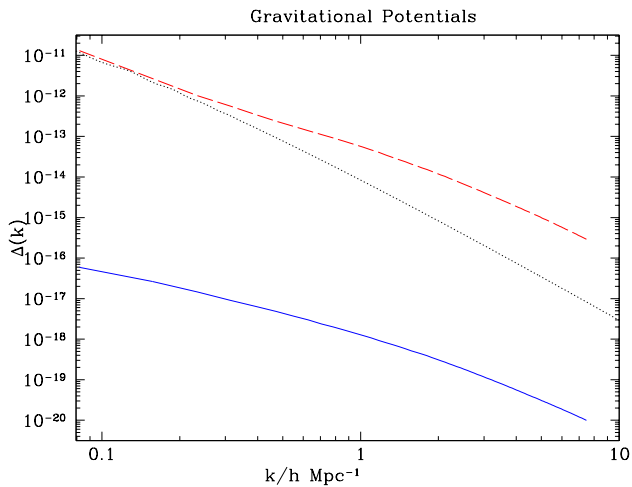}
\end{center}
\caption{The scalar (dashed red line) and vector (solid blue line) gravitational potential power spectra at redshift zero, with the vector potential
calculated using the momentum field method. The linear theory scalar potential is shown for comparison (dotted black line).}
\label{fig_deltavz0}
\end{figure}

In figures \ref{fig_gravz0} and \ref{fig_deltavz0}, we show the power spectrum of the post-Friedmann vector potential as well as the standard Newtonian
scalar potential, at $z=0$, for the curl and momentum field methods of extraction respectively. As expected, both methods show that the scalar potential
is small over all scales and the vector potential is subdominant. There is a quantitative difference between the two methods on the largest scales, but
this difference is not sufficient to alter the expected qualitative behaviour. Notably, the effect of the vector potential on weak-lensing power spectra,
as examined in \cite{1403.4947}, will remain negligible, regardless of which method is used to calculate the vector potential. We have been unable to
determine the reason for this discrepancy and it is unclear to us which method should, a priori, be expected to be more accurate.

\begin{figure}
\begin{center}
\includegraphics[width=2.7in,angle=270]{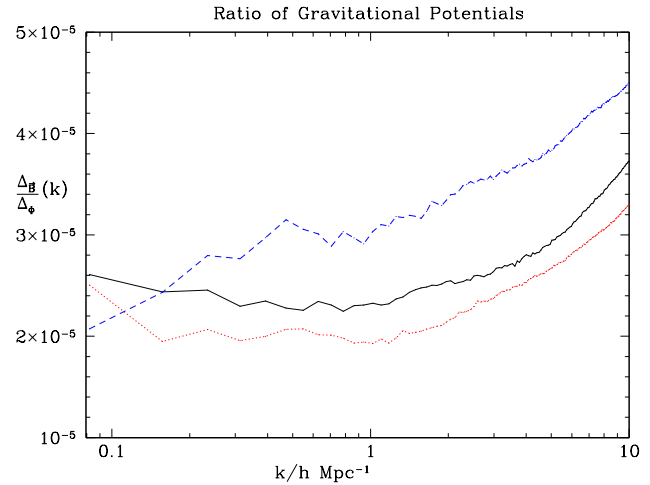}
\end{center}
\caption{The ratio at redshift zero between the vector potential, calculated using the curl method, and the scalar potential. The three curves
show the ratio for the three realisations of the HR simulations.}
\label{fig_potratioz0_err}
\end{figure}

\begin{figure}
\begin{center}
\includegraphics[width=2.7in,angle=270]{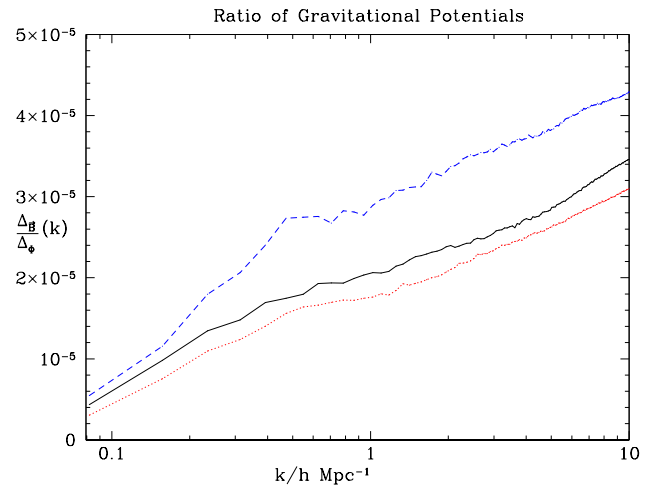}
\end{center}
\caption{The ratio at redshift zero between the vector potential, calculated using the momentum field method, and the scalar potential. The three curves
show the ratio for the three realisations of the HR simulations.}
\label{fig_deltavratioz0_err}
\end{figure}

In figures \ref{fig_potratioz0_err} and \ref{fig_deltavratioz0_err}, we show the ratio between the power spectra of the vector and scalar gravitational
potentials at redshift zero, for the two methods of extracting the vector potential. We plot the ratios for all three individual realisations of the HR
simulations. For the curl method, as shown in \cite{Bruni:2013mua}, this ratio is approximately $2.5\times10^{-5}$. This ratio does not vary significantly
over the range of scales considered, although there is a slight increase towards smaller scales. However, for the momentum field method, the ratio is not
approximately constant due to the decreased power on large scales. We will compare this behaviour to the analytic second-order perturbative behaviour shortly,
here we just note that the curl method produces qualitative behaviour that is closer to the analytic prediction.

\begin{figure}
\begin{center}
\includegraphics[width=2.7in,angle=270]{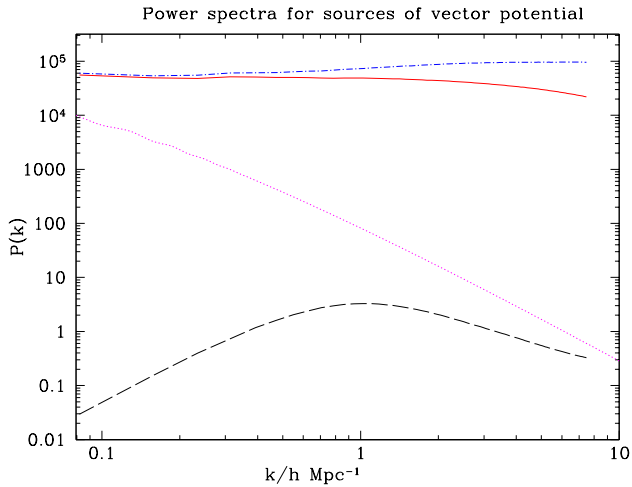}
\end{center}
\caption{The power spectra of the three source terms for the vector potential in equation (\ref{eqn_threecomp}), the vorticity (dashed black line),
$\nabla \delta \times \vec{v}$ (dot-dashed blue line) and $\delta \nabla \times \vec{v}$ (solid red line). The power spectra plotted here are given
by $P(k)/\left(f^2{\cal H}^2(2\pi)^3\right)$, such that the power spectrum of the velocity divergence agrees with the density power spectrum on linear
scales and ensuring that all of the power spectra have the same units, following \protect\cite{pueb}. The linear matter power spectrum is shown as a dotted
magenta line for comparison.}
\label{fig_compz0}
\end{figure}

In figure \ref{fig_compz0}, we show the power spectra of the three sources of the vector potential using the curl method, see equation (\ref{eqn_threecomp}).
The power spectra plotted here are given by $P(k)/\left(f^2{\cal H}^2(2\pi)^3\right)$, where ${\cal H}$ is the conformal time Hubble constant and
$f=d \ln D/d \ln a$ is the logarithmic derivative of the linear growth factor $D$.  These units are chosen such that the power spectrum of the velocity
divergence agrees with the density power spectrum on linear scales and have the same units as the matter power spectrum, following \cite{pueb}. The
vorticity, although often ignored in perturbation theory, is the only one of these three quantities that is linear in perturbations. This figure shows
that it is negligible compared to the other two components, so the vector potential is being predominantly generated by non-linear effects.

Since this vector potential is the first correction to Newtonian theory, this calculation is the first quantitative check of the relationship between
Newtonian simulations and GR on fully non-linear scales. The small magnitude of the vector potential suggests that running Newtonian simulations is
sufficiently accurate for cosmological purposes, whereas a larger calculated value for the vector potential would suggest that the approximations taken
in deriving the fully non-linear Newtonian equations do not hold sufficiently well. As far as relating Newtonian and relativistic cosmologies goes, in
the language of \cite{green12}, the smallness of this vector potential allows the use of the abridged dictionary in \cite{chis11}, rather than the
dictionary proposed in \cite{green12}. We note that the analysis here is for a $\Lambda$CDM cosmology, further work is required to determine the
validity of Newtonian simulations in general dark energy cosmologies.

\subsection{Time evolution}
\begin{figure}
\begin{center}
\includegraphics[width=2.7in,angle=270]{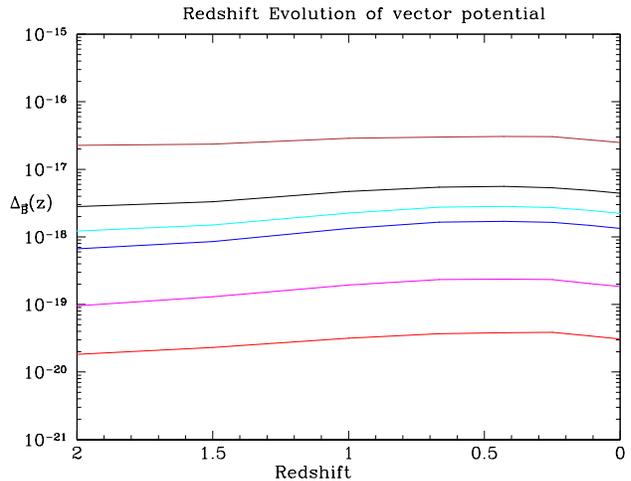}
\end{center}
\caption{The evolution of the vector potential for six different wavenumbers. From top to bottom, these are $k=0.23h$Mpc$^{-1}$ (brown), $k=0.55h$Mpc$^{-1}$ (black),
$k=0.79h$Mpc$^{-1}$ (cyan), $k=1.01h$Mpc$^{-1}$ (blue), $k=2.51h$Mpc$^{-1}$(magenta) and $k=5.03h$Mpc$^{-1}$ (red).}
\label{fig_gravredshifts}
\end{figure}

In this section we will examine the time evolution of the vector potential, and its ratio to the scalar potential, for the redshifts listed in table
\ref{table_redshifts}. The vector potential is this section has been computed using the curl method. In figure \ref{fig_gravredshifts}, we plot the
ratio of the vector potential to the scalar potential as a function of redshift. The different curves in this plot show the evolution for different
wavenumbers. We can see that individual k-modes do not exhibit significant growth over time, although the more non-linear
scales do exhibit slightly more variation in time. Similarly to the scalar gravitational potential, the vector potential at a fixed scale is not monotonic
over time on non-linear scales.

\begin{figure}
\begin{center}
\includegraphics[width=2.7in,angle=270]{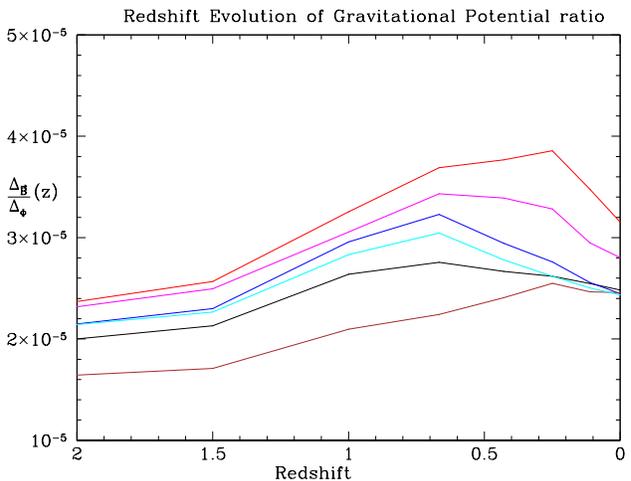}
\end{center}
\caption{The ratio of the vector potential to the scalar potential plotted for six different wavenumbers. From bottom to top (at redshift=1), these are
$k=0.23h$Mpc$^{-1}$ (brown), $k=0.55h$Mpc$^{-1}$ (black), $k=0.79h$Mpc$^{-1}$ (cyan), $k=1.01h$Mpc$^{-1}$ (blue), $k=2.51h$Mpc$^{-1}$(magenta) and
$k=5.03h$Mpc$^{-1}$ (red).}
\label{fig_ratioredshifts}
\end{figure}

In figure \ref{fig_ratioredshifts}, we plot the ratio of the vector potential to the scalar potential as a function of redshift. The different curves
in this plot show the same wavenumbers as in figure \ref{fig_gravredshifts}. The ratio stays fairly constant over time, varying by less than a factor of
two for a given scale. Across the entire range of times and scales under consideration, the ratio varies by less than a factor of 4. The ratio between
the gravitational potentials is also not monotonic over the redshift range under consideration for a given scale.

\subsubsection{Comparison to perturbative calculation}
In \cite{0812.1349}, an analytic calculation of the vector potential was performed using perturbation theory. As a perturbative analysis, it is unclear how
large a value of $k$ this calculation should be extended to. Here we will assume it is valid on all of the scales of overlap between this method and ours.

For the curl method of computing the vector power spectrum, there is similar qualitative behaviour between the two methods, with the ratio of the power
spectra of the vector and scalar potentials being fairly constant and of order $10^{-5}$ in both methods. The difference between the two methods being
that the ratio in \cite{0812.1349} is between the vector and the linear theory scalar potential, whereas the our ratio is between the vector and the
fully non-linear scalar potential. This means that despite this similar qualitative behaviour, the power spectrum of the vector potential in
\cite{0812.1349} underestimates the fully non-linear value on these scales by up to two orders of magnitude, the same factor by which the linear theory
scalar potential power spectrum underestimates the power spectrum of the fully non-linear scalar potential.

The momentum field method of calculating the vector power spectrum results in less-similar qualitative behaviour. It is unclear how well the gravitomagnetic
potential would be expected to match the perturbative prediction on these scales as the velocity field differs from the linear theory at larger scales than
the density.

The power spectrum of the perturbative vector potential is given in \cite{0812.1349} as 
\begin{equation}
 {\cal P}_s(k)=\left(\frac{2 \Delta_{{\cal R}}}{5g_{\infty}}\right)^4\left(\frac{3g\left[g'+\cal{H}g\right]}{\Omega_{m}{\cal H}^2}\right)^2k^2\Pi(u^2)
\end{equation}
where ${\cal P}_s$ is the dimensionless power spectrum of the vector potential, $\Delta_{{\cal R}}$ is the primordial power of the curvature
perturbation, $g$ is the growth factor for the scalar potential, $g_{\infty}$ is a normalisation parameter chosen so $g(0)=1$, $\Pi$ is a function
of the transfer function, $\Omega_m$ is the time dependent matter density and ${\cal H}$ is the conformal Hubble constant. The second term in
parentheses contains all of the time dependence of the vector potential power spectrum and essentially acts as the growth factor for the vector
potential. We  have compared this perturbative prediction for the growth factor of the vector potential to the growth measured in the simulations
(see figure 5 in file CC). This shows that the analytic prediction is not the main source of the time evolution of the vector potential.

\section{Conclusion and Discussion}
\label{sec_conc}
In this paper we have presented the post-Friedmann frame-dragging vector potential calculated on non-linear scales from N-body simulations. We have presented
this vector potential at redshift zero, as well as examining its evolution with redshift. We have also presented the tests we have performed in order to
establish the robustness of our result, including tests of simulation parameters and different methods of extracting the source of the vector potential.\\
We have shown that our density, velocity divergence and vorticity spectra are consistent with the literature and show similar behaviour regarding convergence
tests, particularly mass resolution. We do not see the vorticity scaling with the seventh power of the linear growth factor $D_{+}$ \citep{pueb} beyond $z=1$,
however the differences between our results and others' are within the variance between realisations. We have noted a larger variation of the vorticity than
the density and velocity divergence fields between different realisations, a result that does not seem to have been studied in the literature.\\
We have shown that there is no evidence for a systematic dependence of the vector potential spectrum on box size for boxes smaller than 200$h^{-1}$Mpc, or
on mass resolution with mass resolution better than $6\times10^{8}M_{\odot}$. There is also no evidence that the vector potential is sensitive to the softening
length, binning, number of samples (an internal DTFE parameter) or the grid size used in the analysis. There is a reasonable agreement between the different
methods (curl and momentum field) of extracting the vector potential, although there is an unresolved discrepancy between the two methods on the largest
scales. We do however note the importance of the variation of the vector potential between realisations, this issue is discussed more fully in Appendix
\ref{app_real}.\\
Figures \ref{fig_gravz0} and \ref{fig_gravredshifts} comprise the main physical results of this paper, showing the magnitude of the vector potential power
spectrum at redshift zero and its evolution with time respectively. The magnitude of the vector potential power spectrum can also be expressed in terms of
its ratio to the power spectrum of the scalar potential, as shown in figures \ref{fig_potratioz0_err} and \ref{fig_ratioredshifts}. We have shown that the
power spectrum of the vector potential is around $10^5$ times smaller than the power spectrum of the scalar potential, over a range of scales and redshifts.
These values were used in \cite{Bruni:2013mua} and \cite{1403.4947} when examining the observability of the vector potential, showing that it is neglgible for
currently planned weak-lensing surveys. The small magnitude of the vector potential found here is the first quantitative check of the validity of Newtonian
simulations compared to GR on fully non-linear scales and supports the use of Newtonian simulations for computing cosmological observables. In terms of
interpreting the simulations, the small value of this vector potential seems to justify the use of the abridged dictionary in \cite{chis11}, rather than
the dictionary proposed in \cite{green12}, for relating GR and Newtonain cosmologies.\\
The work carried out so far considers a $\Lambda$CDM cosmology, so this conclusion may no longer be true for a dark energy or modified gravity cosmology.
The post-Friedmann approach would need to be expanded to include modified Einstein equations and/or a fluid with pressure in order to examine alternative
cosmologies and determine whether the use of Newtonian-type N-body simulations is still valid in those cosmologies. The post-Friedmann expansion has been
applied to $f(R)$ gravity and the vector potential calculated from $f(R)$ simulations in \cite{dan_fr}. The vector potential in $f(R)$ was found to be larger
than in General Relativity. We hope that this, and further extensions to
the work in this paper, will allow us to understand how generic the findings in this paper are, and thus justify one of the most widely used tools in
cosmology, N-body simulations.
Whilst this manuscript was being prepared for submission, \cite{adamek_review} appeared on the ArXiv. Their preliminary results seem to agree with the
results of this work. It will be interesting to perform a more in-depth comparison once the details of their work are available.\\

{\sl Acknowledgements} We thank Marius Cautun for help with the publicly available DTFE code and Hector Gil Marin for provision of, and help with,
a Cloud-in-Cells code. We also thank Marc Manera for useful discussions and technical assistance. Some of the numerical computations were done on
the Sciama High Performance Compute (HPC) cluster which is supported by the Institute of Cosmology and Gravitation (ICG) and the University of
Portsmouth. The rest were undertaken on the COSMOS Shared Memory system at DAMTP, University of Cambridge operated on behalf of the STFC DiRAC HPC
Facility. This equipment is funded by BIS National E-infrastructure capital grant ST/J005673/1 and STFC grants ST/H008586/1, ST/K00333X/1. This work
was supported by STFC grants ST/H002774/1, ST/L005573/1 and ST/K00090X/1.

\appendix
\section{Vector power spectra}
\label{app_vec}
We will be dealing with vector quantities, for which there are different ways to define the power spectrum. Our power spectrum for a generic vector
$\vec{v}$ is defined as
\begin{equation}
\langle \tilde {\vec v}(\vec k) \ \cdot \tilde {\vec v} ^*(\vec {k^{'}}) \rangle=(2\pi)^3 \delta^3(\vec k -\vec {k^{'}})P_{\vec v}(k)
\end{equation}
Note that for a divergenceless vector, such as ${\vec B}^N$, $k^2 P_{{\vec B}^N}(k)=P_{\nabla \times {\vec B}^N}(k)$. With our Fourier transform convention,
the dimensionless power spectrum for a field X is given by $\Delta_{X}=k^3 P_{X}(k)/2\pi^2$. All plots of the power spectrum of the vector potential show the
dimensionless power spectrum.

Using equation (\ref{eqn_curl})
\begin{eqnarray}
 &&\hspace{-1.5cm}\langle \widetilde{\nabla \times {\vec B}^N(\vec k)} \ \cdot\widetilde{ \nabla \times {\vec B}^{N*}(\vec {k^{'}})} \rangle=
 \left(\frac{16\pi G \rho_b a^2}{k^2} \right)^2 \nonumber\\
 &&\hspace{-1.5cm}\langle\left[ \widetilde{\left(\nabla \delta\right) \times \vec{v}} +
 \widetilde{\left(1+\delta \right)\nabla \times \vec{v}}\right]\cdot\left[\widetilde{\left(\nabla \delta\right) \times \vec{v}} +
 \widetilde{\left(1+\delta \right)\nabla \times \vec{v}}\right]^*\rangle
\end{eqnarray}
{\footnotesize
\begin{eqnarray}
&&\hspace{-1.5cm}\langle \widetilde{\nabla \times {\vec B}^N(\vec k)} \ \cdot \widetilde{\nabla \times {\vec B}^{N*}(\vec {k^{'}}) }\rangle=
\left(\frac{16\pi G \rho_b a^2}{k^2} \right)^2\nonumber\\
&&\hspace{-1.5cm}\left(\hspace{-0.1cm}
\begin{array}{ccc}
\hspace{-0.2cm}\langle\widetilde{\left[\nabla \delta \times \vec{v}\right]}\cdot \widetilde{\left[\nabla \delta \times \vec{v}\right]}^*\rangle
&\hspace{-0.2cm}+\langle\widetilde{\left[\nabla \delta \times \vec{v}\right]}\cdot \widetilde{\left[\delta \nabla \times \vec{v}\right]}^*\rangle
&\hspace{-0.2cm}+ \langle\widetilde{\left[\nabla \delta \times \vec{v}\right]}\cdot \widetilde{\left[ \nabla \times \vec{v}\right]}^*\rangle\\
\hspace{-0.2cm}+\langle\widetilde{\left[\delta \nabla \times \vec{v}\right]}\cdot \widetilde{\left[\nabla \delta \times \vec{v}\right]}^*\rangle
&\hspace{-0.2cm}+\langle\widetilde{\left[\delta \nabla \times \vec{v}\right]}\cdot \widetilde{\left[\delta \nabla \times \vec{v}\right]}^*\rangle
&\hspace{-0.2cm}+ \langle\widetilde{\left[\delta \nabla \times \vec{v}\right]}\cdot \widetilde{\left[ \nabla \times \vec{v}\right]}^*\rangle\\
\hspace{-0.2cm}+\langle\widetilde{\left[\nabla \times \vec{v}\right]}\cdot \widetilde{\left[\nabla \delta \times \vec{v}\right]}^*\rangle
&\hspace{-0.2cm}+\langle\widetilde{\left[\nabla \times \vec{v}\right]}\cdot \widetilde{\left[\delta \nabla \times \vec{v}\right]}^*\rangle
&\hspace{-0.2cm}+ \langle\widetilde{\left[\nabla \times \vec{v}\right]}\cdot \widetilde{\left[ \nabla \times \vec{v}\right]}^*\rangle
\end{array}\hspace{-0.1cm}
\right)\nonumber
\end{eqnarray}
}
\normalsize

Noting that $A\cdot B^*=(A^*\cdot B)^*$,
\begin{eqnarray}
&&\hspace{-1.5cm} \langle\widetilde{\left[\nabla \times \vec{v}\right]}\cdot \widetilde{\left[\nabla \delta \times \vec{v}\right]}^*\rangle+
\langle\widetilde{\left[\nabla \delta \times \vec{v}\right]}\cdot \widetilde{\left[\nabla \times \vec{v}\right]}\rangle^*=\nonumber\\
&&\hspace{-1.5cm}2 \rm{re}\left( \langle\widetilde{(\nabla \times \vec{v})}\cdot \widetilde{(\nabla \delta \times \vec{v})}\rangle\right)
\equiv(2\pi)^3 \delta^3(\vec k -\vec {k^{'}})P_{\left(\nabla \delta \times \vec{v}\right)\left(\nabla \times \vec{v}\right)}(k) 
\end{eqnarray}

And therefore the dimensionless power spectrum for the vector potential is given by

\begin{equation}
 \Delta_{{\vec B}^N}(k)=\left(\frac{16\pi G \rho_b a^2 }{k^2}\right)^2\frac{k}{2 \pi^2} P_{\delta v}(k)\rm{,}
 \end{equation}
 where
 \begin{eqnarray}
 &&\hspace{-1.2cm}P_{\delta v}(k)=P_{\nabla \times {\vec v}}(k)+P_{\delta \nabla \times {\vec v}}(k)+P_{(\nabla \delta) \times {\vec v}}(k)\nonumber\\
  &&\hspace{-1.2cm}+P_{\left(\nabla \delta \times \vec{v}\right)\left(\nabla \times \vec{v}\right)}(k)
+P_{\left(\nabla \delta \times \vec{v}\right)\left( \delta\nabla \times \vec{v}\right)}(k)
+P_{\left(\delta\nabla \times \vec{v}\right)\left(\nabla \times \vec{v}\right)}(k) 
\end{eqnarray}

\section{Additional robustness information}
\label{app_robust}
In this Appendix we show the figures referred to in the main text as well as discussing additional robustness and convergence tests that were carried out
in order to establish our result.

\begin{figure}
\begin{center}
\includegraphics[width=2.7in,angle=270]{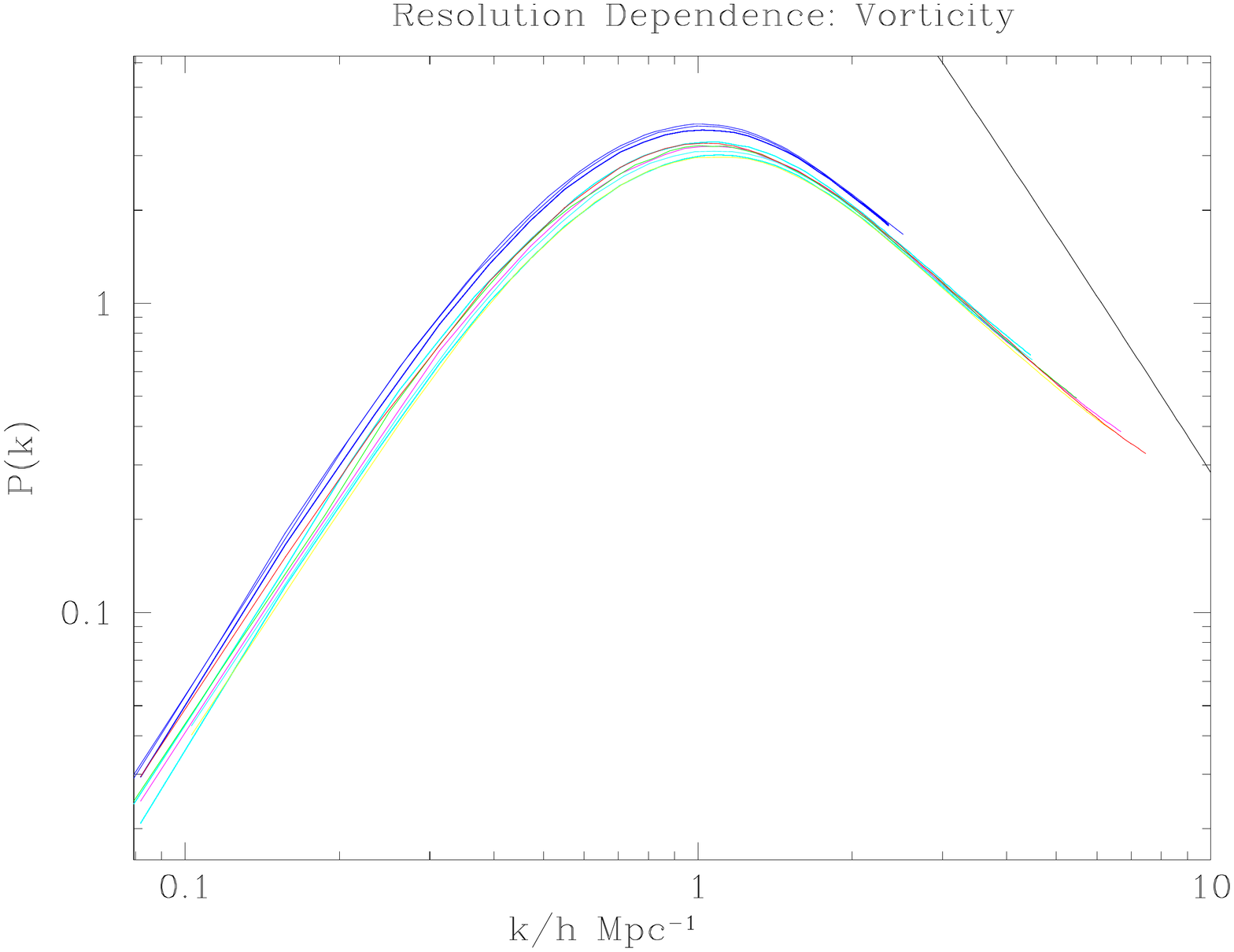}
\end{center}
\caption{The vorticity power spectra extracted from simulations with varying box size and mass resolution. Lines with the same colour share the same mass
resolution (in units of $10^{8}M_{\odot}$: 3.97 (red), 6.26 (magenta), 6.31 (yellow), 7.76 (green), 16.3 (cyan), and 130 (blue). The black curve is the
linear matter power spectrum for comparison.}
\label{fig_vortresn}
\end{figure}

\begin{figure}
\begin{center}
\includegraphics[width=2.7in,angle=270]{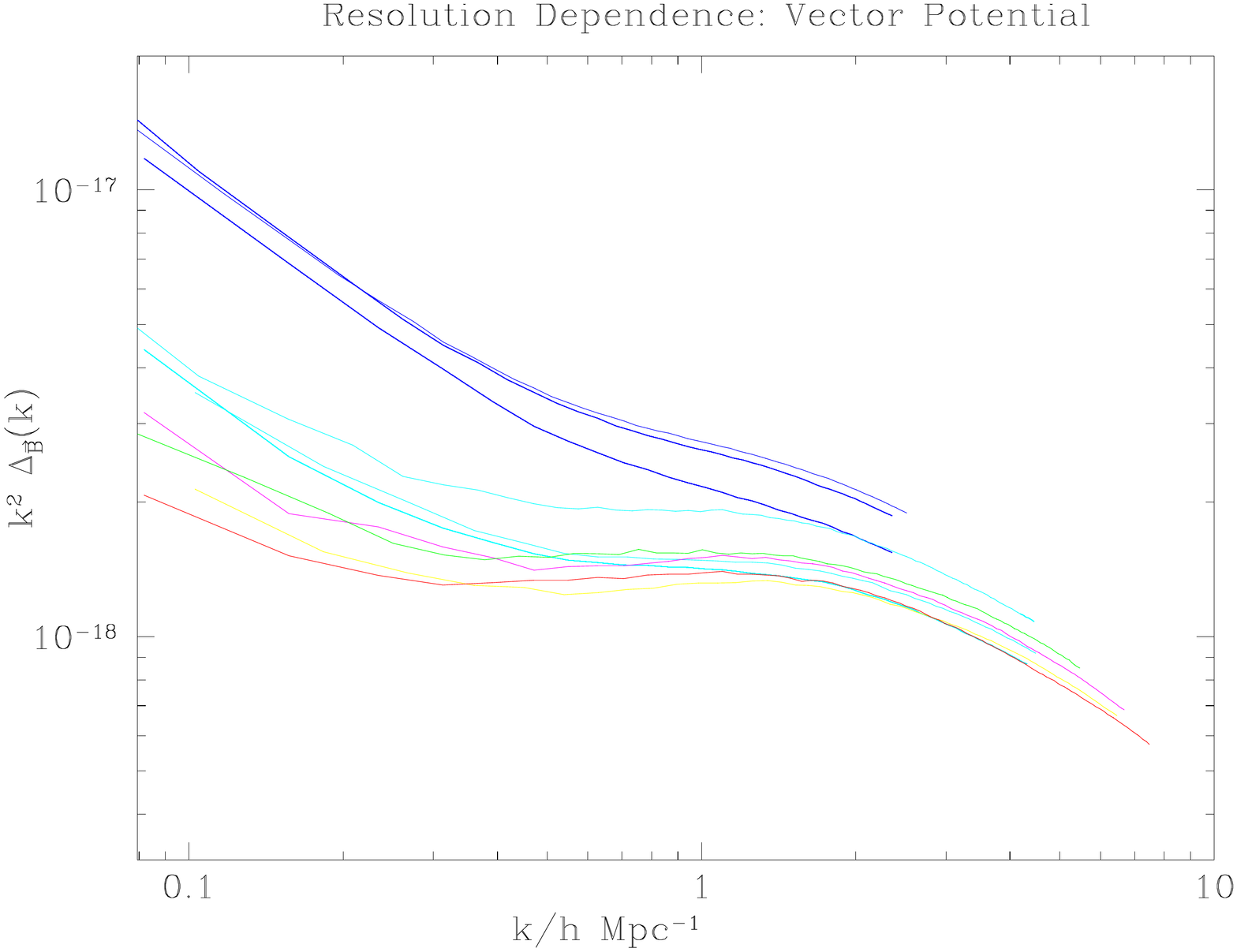}     
\end{center}
\caption{The vector potential power spectra extracted from simulations with varying box size and mass resolution. Lines with the same colour share the same
mass resolution (in units of $10^{8}M_{\odot}$: 3.97 (red), 6.26 (magenta), 6.31 (yellow), 7.76 (green), 16.3 (cyan), and 130 (blue).  The spectra have been
multiplied by $k^2$ in order to better show the variation.}
\label{fig_gravresn}
\end{figure}

\begin{figure}
\begin{center}
\includegraphics[width=2.7in,angle=270]{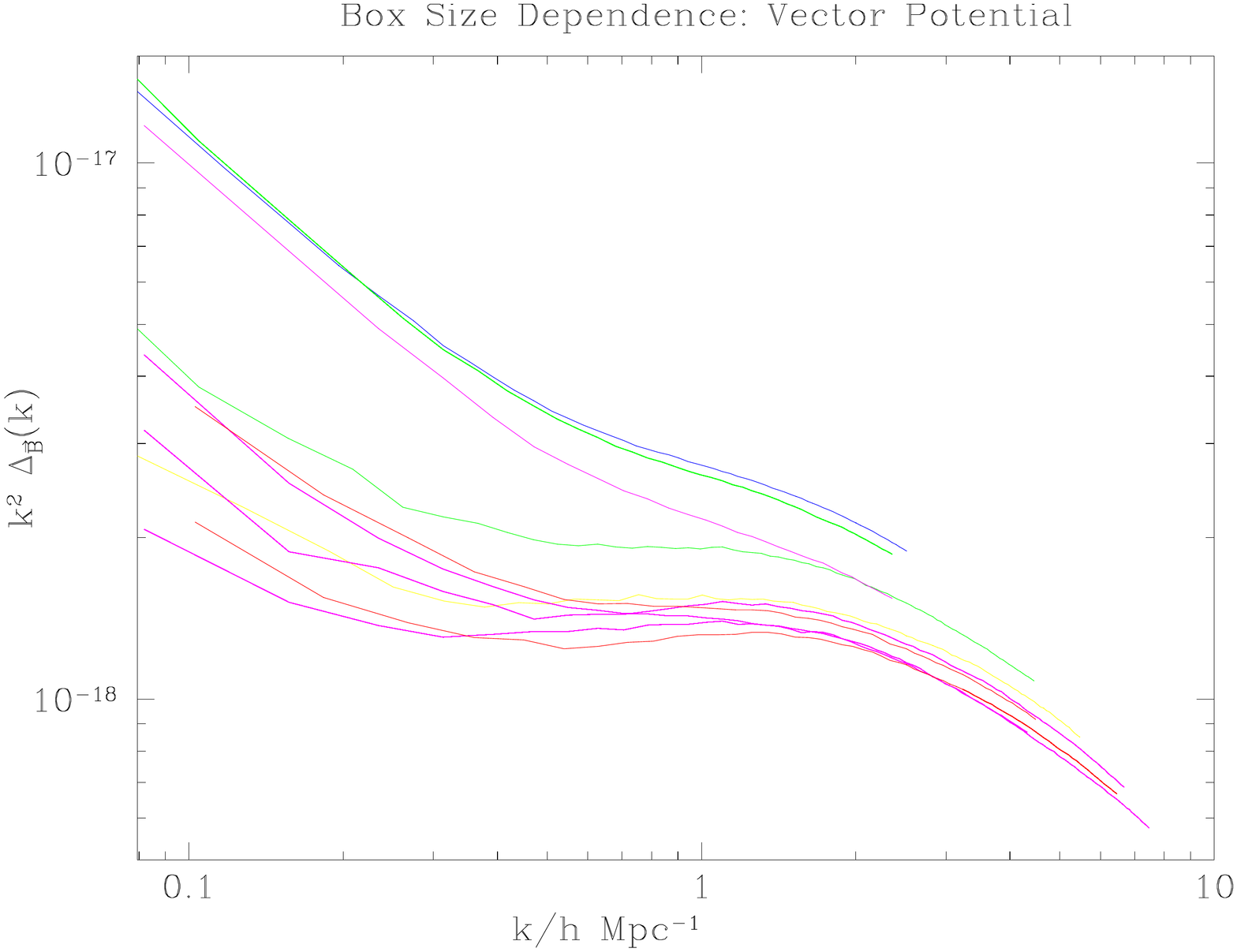}    
\end{center}
\caption{The vector potential power spectra extracted from simulations with varying box size and mass resolution. Lines with the same colour share the same box
size: 140$h^{-1}$Mpc (red), 160$h^{-1}$Mpc (magenta), 200$h^{-1}$Mpc (yellow), 240$h^{-1}$Mpc (green), 320$h^{-1}$Mpc (blue).  The spectra have been multiplied by
$k^2$ in order to better show the variation.}
\label{fig_gravbox}
\end{figure}

\subsection{DTFE parameters}
There are several internal DTFE parameters that are used when computing these fields on a regular grid. We investigate the effects of two of these parameters
here, the grid size and the number of samples that are made in each grid cell, $N_{\rm{samples}}$.

We examined the effect of varying the grid size on the extracted density, velocity divergence and vorticity power spectra. In all cases the agreement is very
good, except on the smallest scales. A discrepancy on this scales is expected due to the change in the resolution of the grid and the effects of the DTFE window
function. However, even on the smallest scales, the discrepancy is small. This can be seen in figures 1, 2 and 3 in file BG, where we show the extracted spectra
from one of the 160$h^{-1}$Mpc simulations with $1024^3$ particles at redshift zero. The different lines show the different grid sizes used: 1024 (blue), 950
(cyan), 850 (green), 750 (magenta) and 640 (red). The black line shows the linear density power spectrum for comparison. For our results plots, we have used
the suggested value $N^3_{\rm{grid}}=N_{\rm{part}}=1024$.

\begin{figure}
\begin{center}
\includegraphics[width=2.7in,angle=270]{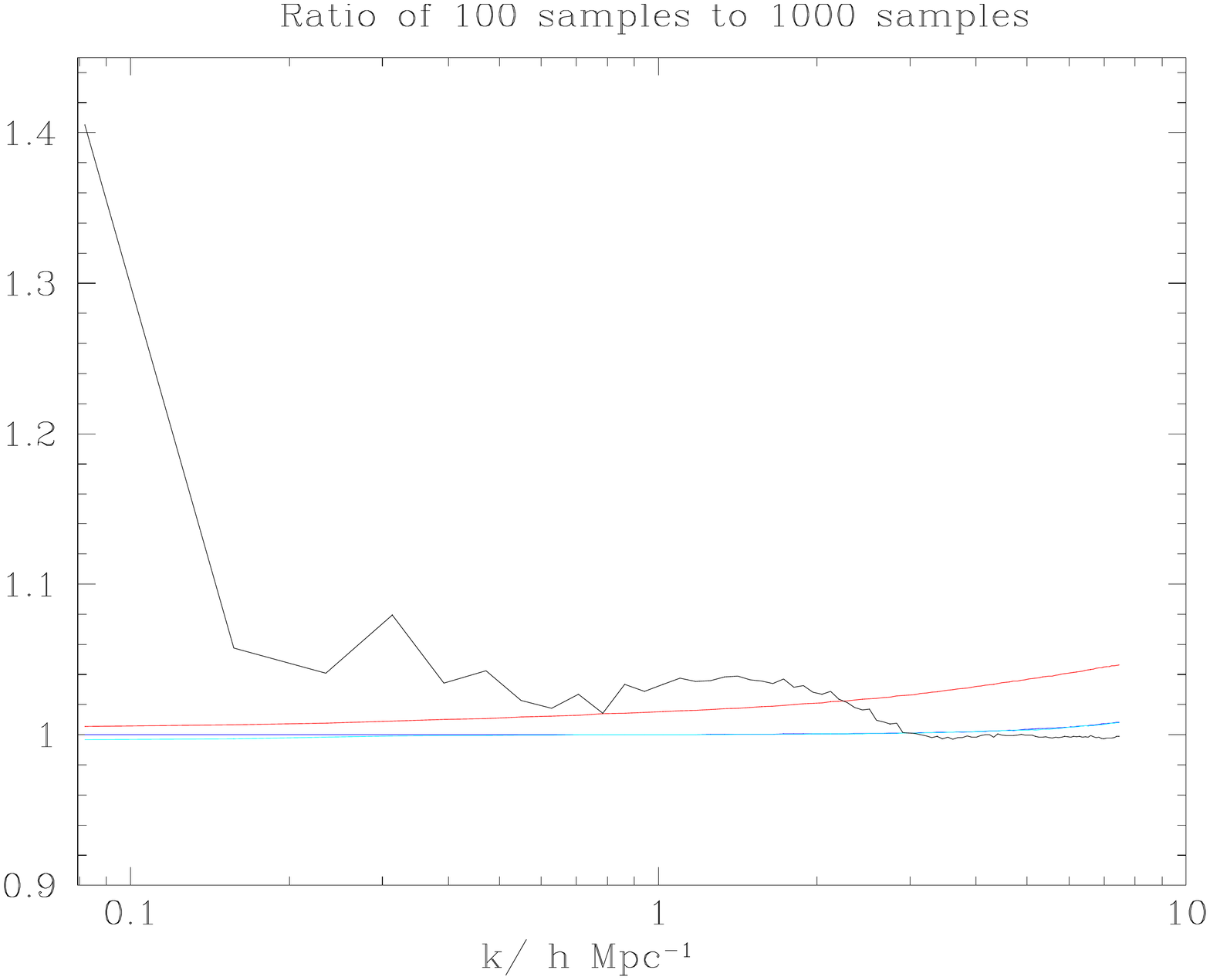}
\end{center}
\caption{The ratios of the power spectra computed with $N_{\rm{samples}}=100$ and $N_{\rm{samples}}=1000$. The ratios shown are for the density (red), velocity
divergence (blue), vorticity (cyan) and vector potential (black).}
\label{fig_samples}
\end{figure}

Our analysis has all been carried out with $N_{\rm{samples}}=100$ points per grid cell, partly due to computing constraints; increasing the number of samples
increases the run time and memory required when analysing a snapshot. However, in figure \ref{fig_samples} we show the effect of increasing $N_{\rm{samples}}$
to 1000 points per grid cell for one of the 160$h^{-1}$ Mpc simulations with $1024^3$ particles. The velocity divergence and vorticity spectra agree very well
between the two different numbers of samples. The density power spectrum shows a deviation that increases towards smaller scales, however is within $5\%$ for
the range of scales under consideration here. The power spectrum of the vector potential shows more deviation, with decreasing deviation for smaller scales.
However, the change in the vector potential is within $10\%$ for every bin after the first and is within $5\%$ for all scales $k>0.3h^{-1}$Mpc.

\subsection{Linear evolution}\label{app_linear}
A further check that can be performed is to examine how the time variation of our extracted density, velocity divergence and vorticity power spectra compares to
the respective linear predictions. For the density and velocity divergence fields, the power spectra evolve as $(D_{+}(z)/D_{+}(z=0))^2$ on the largest scales and
earliest times, as per the linear theory prediction. This prediction becomes increasingly inaccurate for more scales at lower redshifts due to non-linear effects.

\begin{figure}
\begin{center}
\includegraphics[width=2.7in,angle=270]{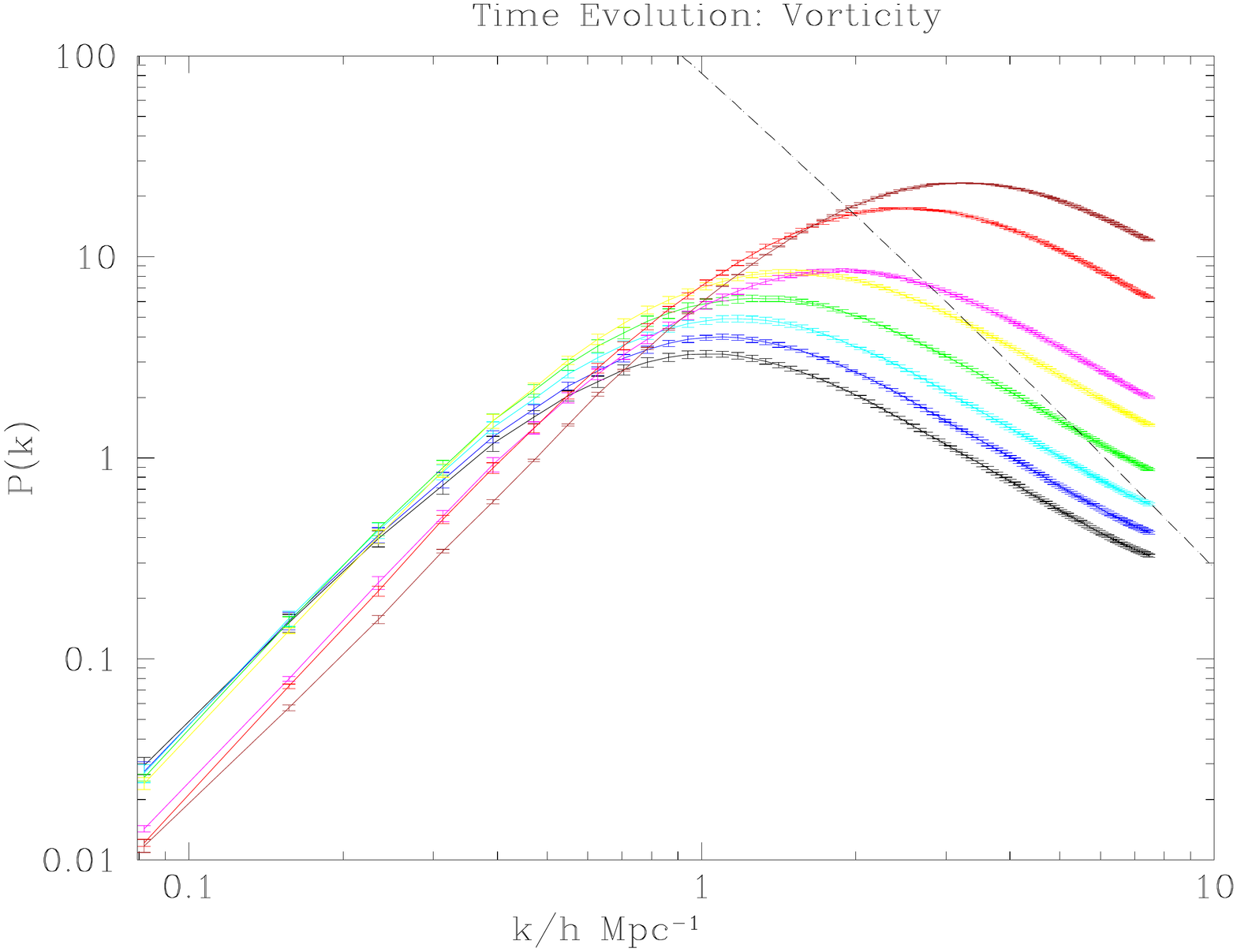}
\end{center}
\caption{The non-linear vorticity power spectrum at selected redshifts, $z=2$ to $z=0$, each divided by the respective linear theory density
growth factor to the seventh power, see \protect\cite{pueb}. See table \ref{table_redshifts} for details and an explanation of the colours. As expected,
the linear theory prediction works well on the largest scales and is generally worse for smaller scales and later times, however the scaling as
the seventh power of the density growth factor seems to break down at earlier times. The error bars on this plot show the standard error on the
mean for each set of realisations.}
\label{fig_vort_lin_evoln_err}
\end{figure}

The time evolution that we found for the vorticity is shown in figure \ref{fig_vort_lin_evoln_err}. In this figure, the power
spectrum at each redshift has been divided by the seventh power of the linear growth factor for that redshift, $(D_{+}(z)/D_{+}(z=0))^7$, as suggested by
\cite{pueb}. In \cite{pueb}, the authors include an approximate analytic derivation of the time evolution of the vorticity power spectrum, finding it to behave
as $f_v^2(z)D_{+}^6(z))$, where $f_v(z)$ is the fraction of the volume that undergoes orbit crossing. Fitting to their simulations, they found
$(D_{+}(z)/D_{+}(z=0))^{7\pm0.3}$ to be the best fit value. The scaling of our vorticity spectrum appears similar to that found in \cite{pueb}. However, in our
simulations this scaling appears to break down for higher redshift, $z\geq1$. We see a smaller vorticity spectrum at these times than expected from the
$(D_{+}(z)/D_{+}(z=0))^7$ scaling. Figure \ref{fig_vort_lin_evoln_err} shows this discrepancy along with the error amongst our simulations. These errors do not
appear sufficiently large to explain the discrepancy. However, it is worth noting that the variation amongst our realisations (see Appendix \ref{app_real}) is
large enough to explain the difference in the time evolution of the vorticity between our simulations and the single high resolution simulation in \cite{pueb}.
The variation between realisations was not considered in \cite{pueb}, however it seems likely that the function $f_v(z)$ varies between realisations. The range
of the scaling of the vorticity with the linear growth factor has an upper value of 7.3 in \cite{pueb}. Using this value
reduces, but does not remove the discrepancy.

\subsection{Comparison with the POWMES density power spectrum}
The density and density gradient power spectra (the latter divided by $k^2$, see the consistency check above) that we have extracted can be compared
to the density field extracted by POWMES \citep{powmes}, a state of the art conventional density power spectrum estimator. For the HR simulations,
the power spectra agree within $10\%$ for $0.2h\rm{Mpc}^{-1} \leq k \leq 7.0 h$Mpc$^{-1}$, see figure 3 in file CC, and within $5\%$ for
the majority of this range. A similar result is seen for the ratio of the DTFE gradient of the density spectrum (divided by $k^2$) to the POWMES
density spectrum, see figure 4 in file CC.

The agreement on the largest scales, in the first 4-5 bins, is affected by the choice of binning. If the number of bins used for the DTFE extraction
is doubled, then the DTFE and POWMES extractions agree much more closely as the bins are then of a more similar size and location. As noted in Appendix
\ref{app_bins}, if we increase the number of bins then the number of $k$ modes contributing to the first few bins is much smaller, so we will continue
to use $N_{\rm{grid}}/4$ bins in our analysis. The agreement between the POWMES and DTFE methods is sufficient to support the robustness of our density
and density gradient spectra.

\subsection{Realisations}
\label{app_real}
In this section we show how the extracted spectra vary amongst realisations. We will illustrate this with the 160$h^{-1}$Mpc $640^3$ particle simulations
for which there are 8 realisations. In all cases we consider the variation at redshift zero.

We examined the variation amongst realisations for the density field, using both the DTFE code and POWMES, and also the velocity divergence. These all showed
the expected variation, namely that cosmic variance causes a difference between the realisations on the largest scales in each box, but this difference is
much reduced on smaller scales. The variance between realisations for the density field was very similar for the two methods of extracting the density field.

\begin{figure}
\begin{center}
\includegraphics[width=2.7in,angle=270]{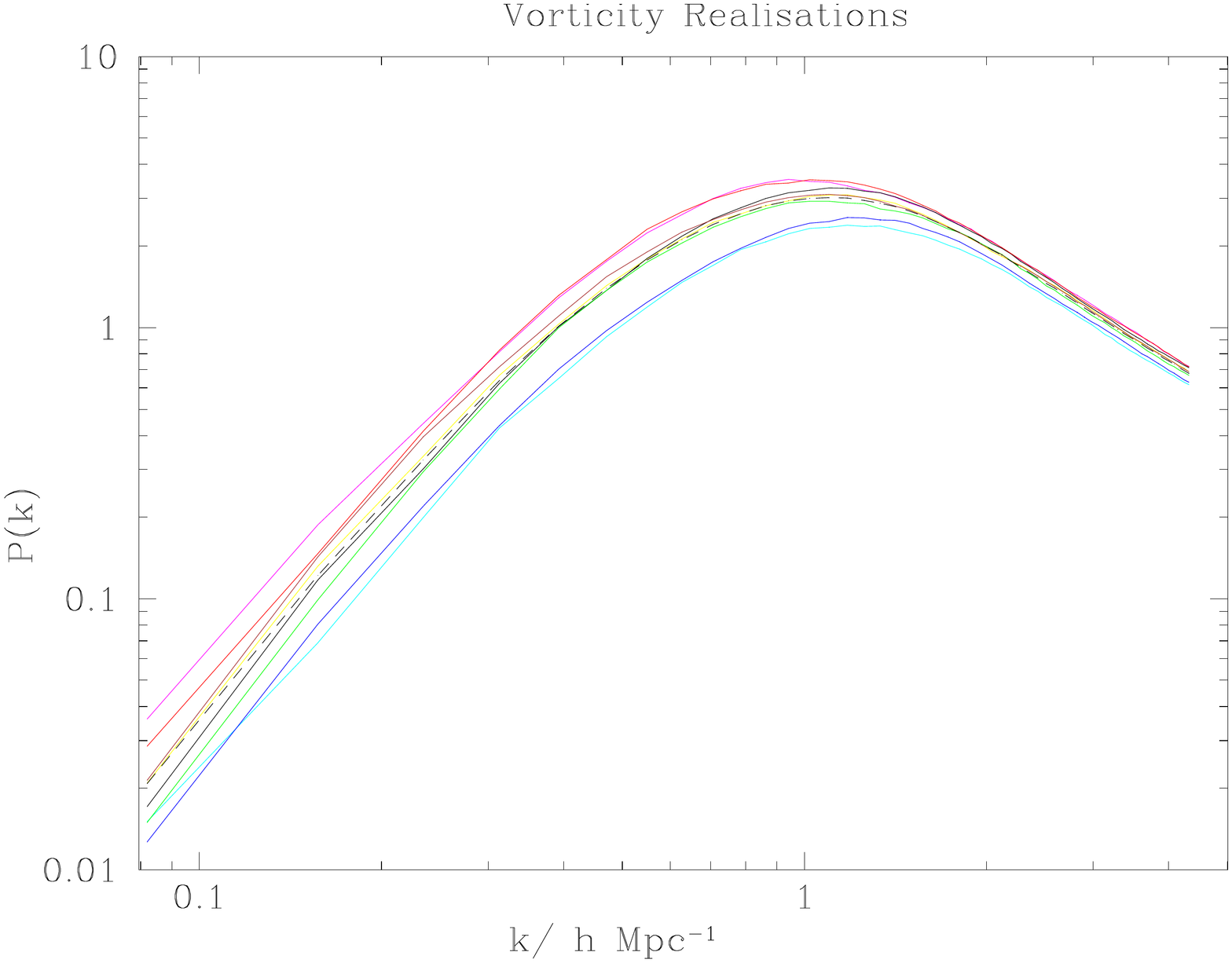}
\end{center}
\caption{The vorticity power spectra as extracted from the 8 160$h^{-1}$Mpc simulations with $640^3$ particles. The dashed black line denotes the average
of the 8 simulations.}
\label{fig_vort640realisations}
\end{figure}

In figure \ref{fig_vort640realisations} we show the variation of the vorticity field amongst realisations. This plot shows that the variation amongst realisations
is greater for the vorticity than for the density. On smaller scales, the variation amongst realisations of the vorticity is less than on large scales, but still
greater than for the density field. We are not aware of this being previously noted in the literature, and the works \citep{pueb,1308.0886,1404.2280}
that we compare our vorticity spectrum to in the main text do not have multiple realisations in order to have seen this effect.

\begin{figure}
\begin{center}
\includegraphics[width=2.7in,angle=270]{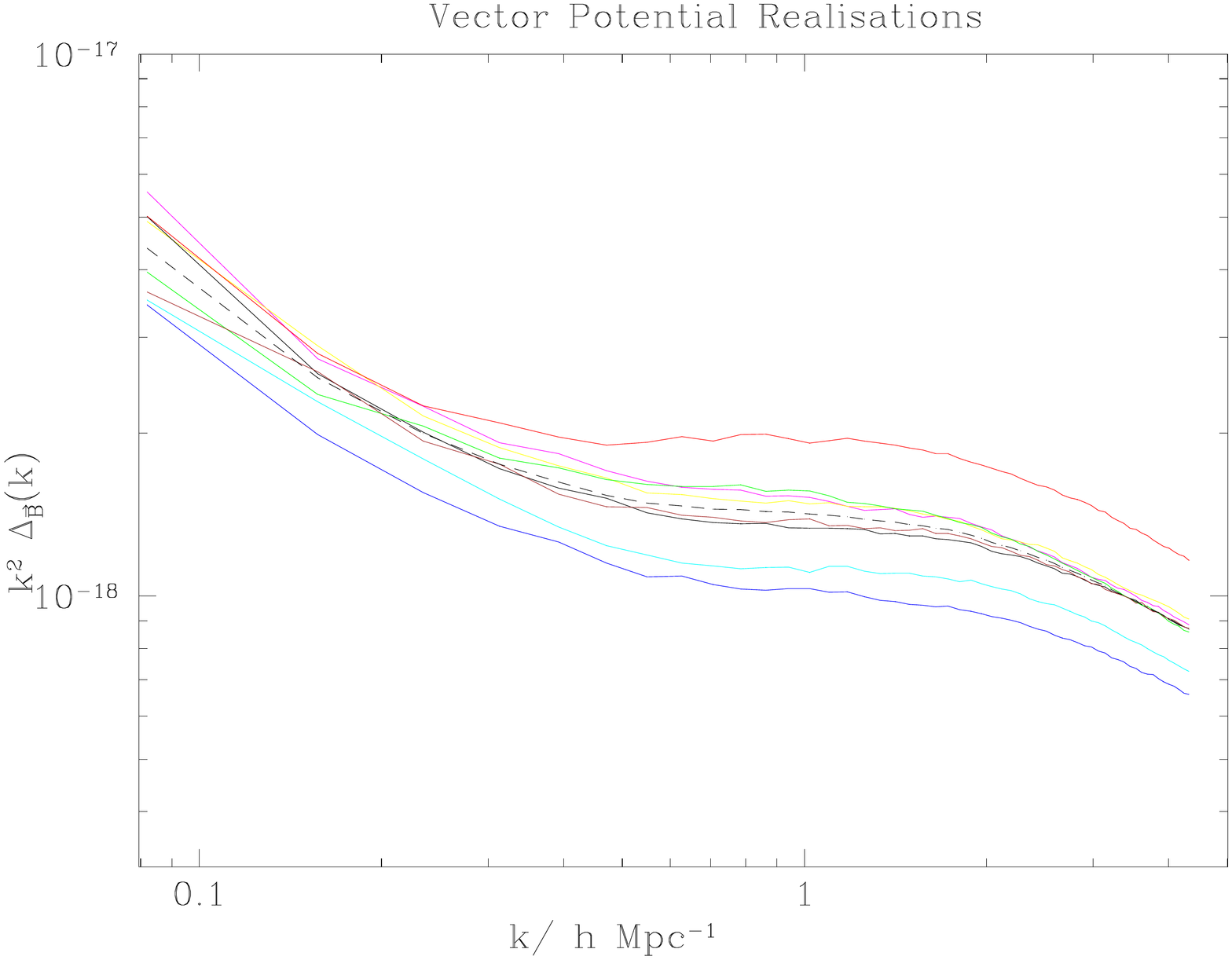}    
\end{center}
\caption{The vector potential power spectra as extracted from the 8 160$h^{-1}$Mpc simulations with $640^3$ particles. The dashed black line denotes the
average of the 8 simulations. The spectra have been multiplied by $k^2$ in order to better show the variation.}
\label{fig_grav640realisations}
\end{figure}

In figure \ref{fig_grav640realisations}, we show the variation amongst realisations of the vector potential. On large
scales, the variation between realisations is very similar to that between the vorticity spectra. However, the variation does not appear to reduce on smaller
scales. According to perturbative results \citep{0709.1619,0812.1349}, the vector is generated most efficiently by coupling between two different $k$ modes,
particularly if one of them is entering the horizon. Given the similar qualitative behaviour of the fully non-linear vector potential, it is reasonable to
assume that this is also generated by coupling between large scale modes and small scale modes. Thus, the large scale variance between realisations will be
affecting the vector power spectrum on smaller scales, resulting in the variance between realisations not decreasing on small scales.

\begin{figure}
\begin{center}
\includegraphics[width=2.7in,angle=270]{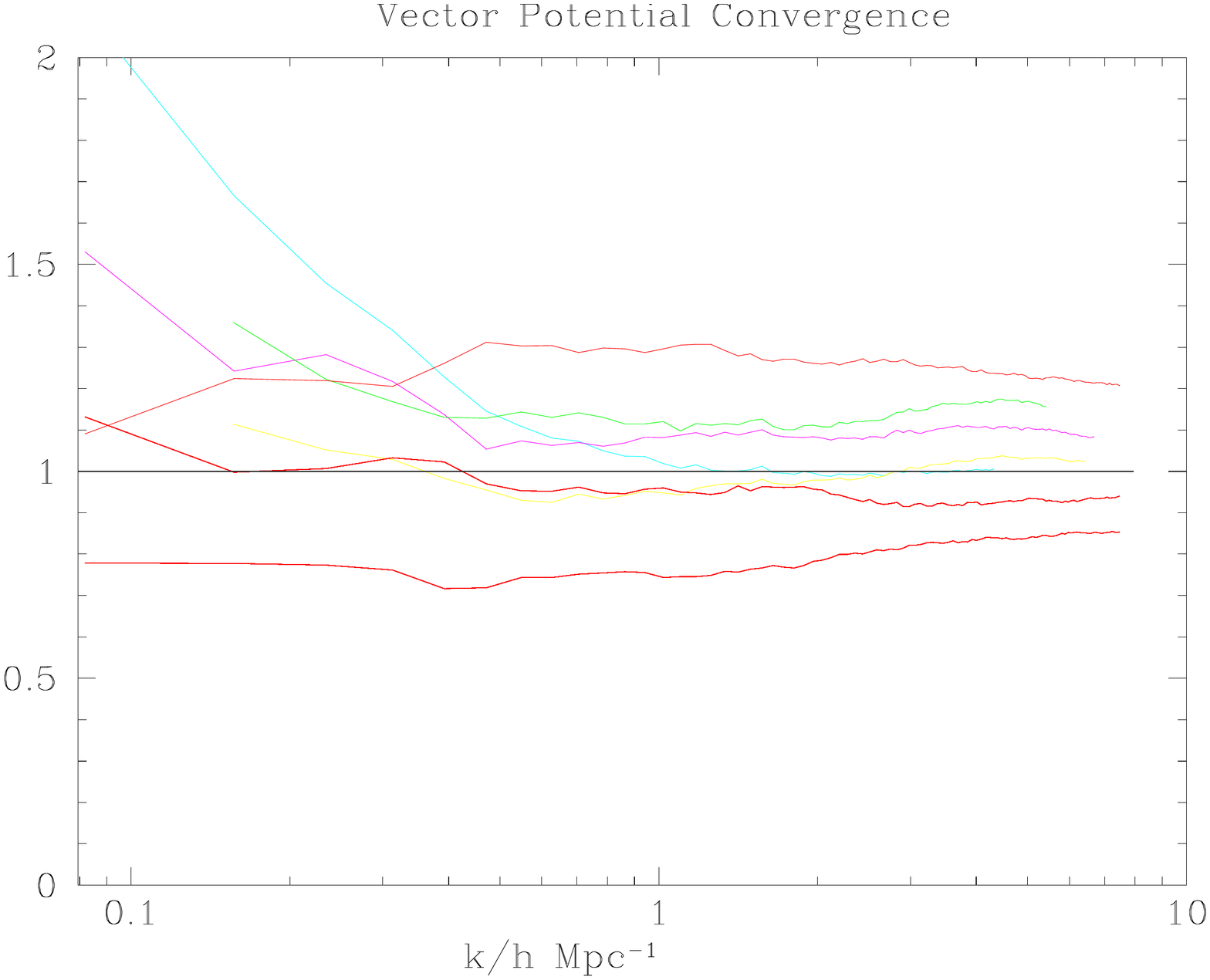}  
\end{center}
\caption{The vector potential power spectra from different simulations, divided by the average vector potential from the three 160$h^{-1}$Mpc simulations
with $1024^3$ particles. The three red curves show the vector potential from the three realisations of the 160$h^{-1}$Mpc simulations with $1024^3$
particles. The cyan and magenta curves show the vector potential from the average of the 160$h^{-1}$Mpc simulations with $640^3$ and $880^3$ particles
respectively. The yellow curve shows the average of the 140$h^{-1}$Mpc simulations with $768^3$ particles and the green curves shows the average of the
200$h^{-1}$Mpc simulations with $1024^3$ particles. Note that the variation between the high resolution simulations is greater than the variation between
the average values from simulations with different parameters.}
\label{fig_gravratiovsreal}
\end{figure}

In figure \ref{fig_gravratiovsreal}, we show how the value of the vector potential from the individual realisations of the HR simulations compares to the average
over realisations of simulations with different parameters. Note that the variation between the HR realisations is greater than the variation between the
average over realisations for different simulation parameters.

As mentioned above, the increased variance between realisations may be an unavoidable feature of the vector potential. As such, this represents the dominant
source of error in calculating the vector potential, as long as the simulation parameters are sufficiently good. If an observational test of the vector
potential was found, then many more realisations than the number carried out for this paper would be required, in order to more carefully investigate this
effect and determine more precisely what the observational prediction would be for a $\Lambda$CDM cosmology.

\subsection{Softening length}
\label{app_soft}
\begin{figure}
\begin{center}
\includegraphics[width=2.7in,angle=270]{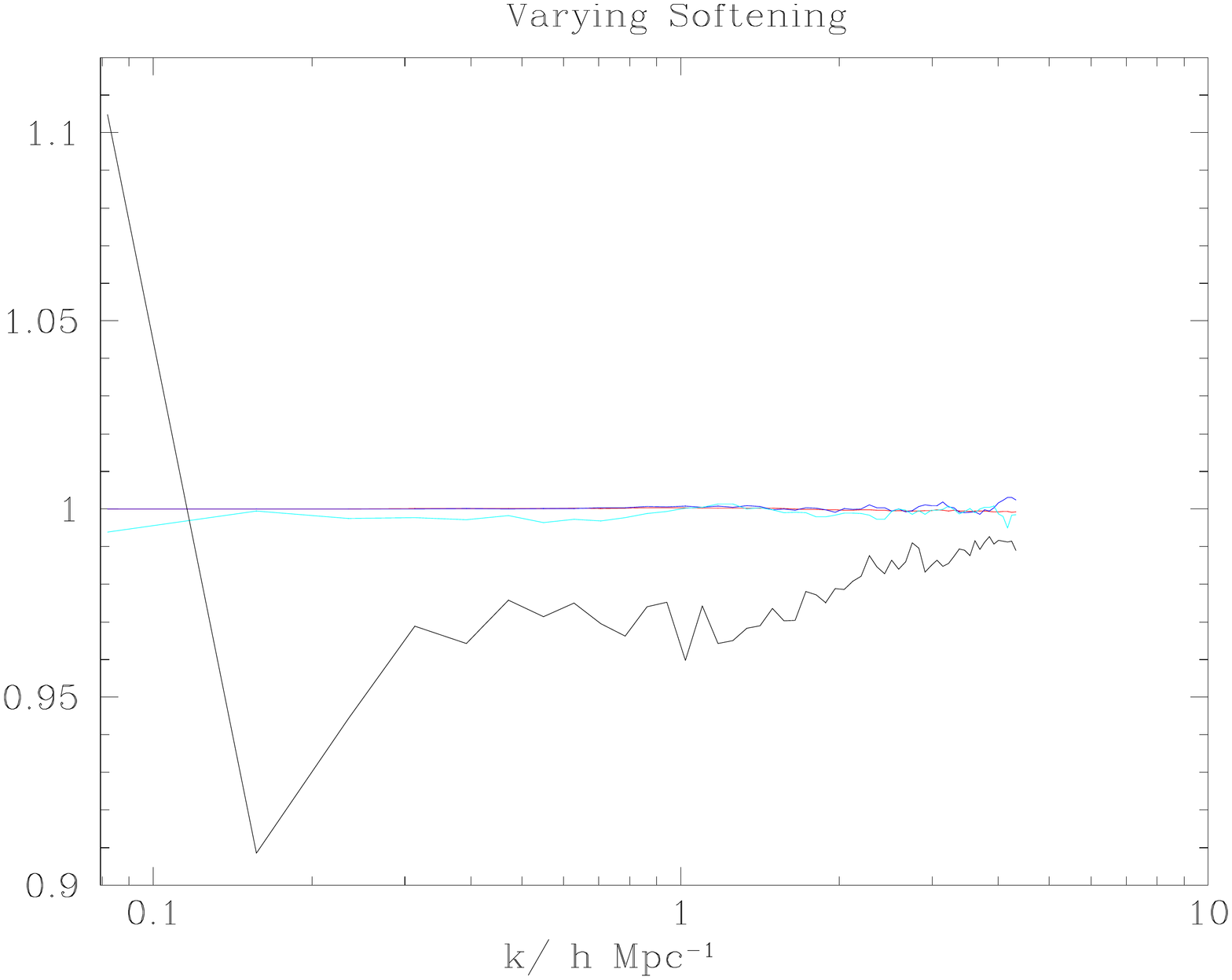}
\end{center}
\caption{The ratio between the power spectra extracted at redshift zero for the same initial conditions run with two different softening lengths. The
different spectra plotted are density (red), velocity divergence (blue), vorticity (cyan) and the vector potential (black).}
\label{fig_softening160}
\end{figure}

In this paper we have chosen our softening lengths following \cite{pueb} in order to compare to their results. In figure \ref{fig_softening160}, we show
how a 160$h^{-1}$Mpc simulation with $640^3$ particles and the same initial conditions varies if the softening length changes from 6.5kpc to 5kpc. This
is a 20$\%$ change in the softening length. The variation between the density, velocity divergence and vorticity spectra is very small for this change.
The power spectrum of the vector potential varies more, but is within 5$\%$ of the value for nearly the entire range under consideration. Since this
5$\%$ variation is significantly smaller than the 20$\%$ variation in the softening length, we do not think the choice of softening length significantly
impacts our results for a sensible choice of softening length.

\subsection{Smaller Boxes}
\label{app_smallboxes}
Here we examine some additional plots that demonstrate some of the comments made in the main text. We ran a set
of 8 simulations with $512^3$ particles in an 80$h^{-1}$Mpc box and extracted the power spectra in the same way as from our other simulations.
In figures 3, 5 and 6 in file RB, we show the density, vorticity and vector potential power spectra respectively, colour coded to match box size as
in figure \ref{fig_gravbox}. In addition, the power spectra extracted from the smaller boxes is shown as a dashed black line. It is clear that the spectra
extracted from the smaller 80$h^{-1}$Mpc boxes are systematically smaller, irrespective of any other dependence on box size and resolution.
The effect of using (too) small boxes when running N-body simulations has been examined in the past, see e.g. \cite{9408029,0410373,0601320}. 
A suggestion in \cite{0601320} is that it is important that the ratio between the box size and the scale of non-linearity is sufficiently large. As a result, for our
simulations, the smallest boxes we have run that we consider trustworthy are the 140$h^{-1}$Mpc simulations. It remains to be seen whether smaller boxes,
such as the 100$h^{-1}$Mpc simulations used in \cite{1308.0886} are a robust source of spectra such as the density and vorticity.

\subsection{Number of bins}
\label{app_bins}
We considered the effect on our extracted power spectra of varying the number of bins. As expected, increasing the number of bins increases the noise
of the power spectra and there is no systematic deviation. Our results from varying the number of bins on the density, velocity divergence and vorticity
power spectra are shown in figures 4-6 in file GB. In each of these plots, the 256 bins used for the analysis in this paper is shown by the black line,
the blue lines denote the use of 512 bins and the red lines are for 1024 bins. We have used 256 bins for our analysis to ensure that the low $k$ bins
contain a sufficient number of $k$-modes. For the 256 bins, the first two $k$ bins contain 58 and 218 $k$-modes respectively, whereas these numbers are
12 and 41 for the corresponding bins when 1024 bins are used. Note that, as mentioned in the POWMES section, the variation between the 256 bin and 512
bin power spectra is similar to the variation between the POWMES method and the DTFE method using 256 bins. This is due to the number and location of bins
in the POWMES method being very similar to the DTFE method with 512 bins.

In addition, figure 7 in file GB shows the variation of the vector potential power spectrum with the number of bins. Again, the change in the number
of bins is negligible. In this plot, the dashed lines show the power spectra computed with the extra factors of k explicitly included whilst summing over
the modulus squared values of the field, see the velocity consistency check for more information. As expected, this change affects things the most in the
largest bins and therefore on large scales and for the smallest number of bins, however it does not affect our results.

\bibliographystyle{mn2e}
\bibliography{portslonger}

\begin{thebibliography}{}

\bibitem[\protect\citeauthoryear{{Adamek}, {Daverio}, {Durrer} \&
  {Kunz}}{{Adamek} et~al.}{2013}]{1308.6524}
{Adamek} J.,  {Daverio} D.,  {Durrer} R.,    {Kunz} M.,  2013, \prd, 88, 103527

\bibitem[\protect\citeauthoryear{{Adamek}, {Durrer} \& {Kunz}}{{Adamek}
  et~al.}{2014}]{adamek_review}
{Adamek} J.,  {Durrer} R.,    {Kunz} M.,  2014, Classical and Quantum Gravity,
  31, 234006

\bibitem[\protect\citeauthoryear{{Bagla} \& {Prasad}}{{Bagla} \&
  {Prasad}}{2006}]{0601320}
{Bagla} J.~S.,  {Prasad} J.,  2006, \mnras, 370, 993

\bibitem[\protect\citeauthoryear{{Bagla} \& {Ray}}{{Bagla} \&
  {Ray}}{2005}]{0410373}
{Bagla} J.~S.,  {Ray} S.,  2005, \mnras, 358, 1076

\bibitem[\protect\citeauthoryear{{Bernardeau} \& {van de
  Weygaert}}{{Bernardeau} \& {van de Weygaert}}{1996}]{dtfe3}
{Bernardeau} F.,  {van de Weygaert} R.,  1996, \mnras, 279, 693

\bibitem[\protect\citeauthoryear{Bruni, Thomas \& Wands}{Bruni
  et~al.}{2014}]{Bruni:2013mua}
Bruni M.,  Thomas D.~B.,    Wands D.,  2014, Phys.Rev., D89, 044010

\bibitem[\protect\citeauthoryear{{Carbone} \& {Matarrese}}{{Carbone} \&
  {Matarrese}}{2005}]{carbone}
{Carbone} C.,  {Matarrese} S.,  2005, \prd, 71, 043508

\bibitem[\protect\citeauthoryear{{Cautun} \& {van de Weygaert}}{{Cautun} \&
  {van de Weygaert}}{2011}]{dtfecode}
{Cautun} M.~C.,  {van de Weygaert} R.,  2011, ArXiv e-prints

\bibitem[\protect\citeauthoryear{{Chandrasekhar}}{{Chandrasekhar}}{1965}]{chand}
{Chandrasekhar} S.,  1965, \apj, 142, 1488

\bibitem[\protect\citeauthoryear{{Chisari} \& {Zaldarriaga}}{{Chisari} \&
  {Zaldarriaga}}{2011}]{chis11}
{Chisari} N.~E.,  {Zaldarriaga} M.,  2011, \prd, 83, 123505

\bibitem[\protect\citeauthoryear{{Colombi}, {Jaffe}, {Novikov} \&
  {Pichon}}{{Colombi} et~al.}{2009}]{powmes}
{Colombi} S.,  {Jaffe} A.,  {Novikov} D.,    {Pichon} C.,  2009, \mnras, 393,
  511

\bibitem[\protect\citeauthoryear{{Crocce}, {Pueblas} \& {Scoccimarro}}{{Crocce}
  et~al.}{2006}]{2lptb}
{Crocce} M.,  {Pueblas} S.,    {Scoccimarro} R.,  2006, \mnras, 373, 369

\bibitem[\protect\citeauthoryear{{Everitt et al}}{{Everitt et al}}{2011}]{gpb}
{Everitt et al} C.~W.~F.,  2011, Physical Review Letters, 106, 221101

\bibitem[\protect\citeauthoryear{{Flender} \& {Schwarz}}{{Flender} \&
  {Schwarz}}{2012}]{1207.2035}
{Flender} S.~F.,  {Schwarz} D.~J.,  2012, \prd, 86, 063527

\bibitem[\protect\citeauthoryear{{Gelb} \& {Bertschinger}}{{Gelb} \&
  {Bertschinger}}{1994}]{9408029}
{Gelb} J.~M.,  {Bertschinger} E.,  1994, \apj, 436, 491

\bibitem[\protect\citeauthoryear{{Green} \& {Wald}}{{Green} \&
  {Wald}}{2012}]{green12}
{Green} S.~R.,  {Wald} R.~M.,  2012, \prd, 85, 063512

\bibitem[\protect\citeauthoryear{{Hahn}, {Angulo} \& {Abel}}{{Hahn}
  et~al.}{2014}]{1404.2280}
{Hahn} O.,  {Angulo} R.~E.,    {Abel} T.,  2014, ArXiv e-prints

\bibitem[\protect\citeauthoryear{{Haugg}, {Hofmann} \& {Kopp}}{{Haugg}
  et~al.}{2012}]{1211.0011}
{Haugg} T.,  {Hofmann} S.,    {Kopp} M.,  2012, ArXiv e-prints

\bibitem[\protect\citeauthoryear{{Hockney} \& {Eastwood}}{{Hockney} \&
  {Eastwood}}{1981}]{cic}
{Hockney} R.~W.,  {Eastwood} J.~W.,  1981, {Computer Simulation Using
  Particles}

\bibitem[\protect\citeauthoryear{{Hui-Ching Lu}, {Ananda}, {Clarkson} \&
  {Maartens}}{{Hui-Ching Lu} et~al.}{2009}]{0812.1349}
{Hui-Ching Lu} T.,  {Ananda} K.,  {Clarkson} C.,    {Maartens} R.,  2009,
  \jcap, 2, 23

\bibitem[\protect\citeauthoryear{{Hwang} \& {Noh}}{{Hwang} \&
  {Noh}}{2013}]{hwangnonlin}
{Hwang} J.-c.,  {Noh} H.,  2013, \jcap, 4, 35

\bibitem[\protect\citeauthoryear{{Hwang}, {Noh} \& {Puetzfeld}}{{Hwang}
  et~al.}{2008}]{hwangpn}
{Hwang} J.-c.,  {Noh} H.,    {Puetzfeld} D.,  2008, \jcap, 3, 10

\bibitem[\protect\citeauthoryear{{Koda}, {Blake}, {Davis}, {Magoulas},
  {Springob}, {Scrimgeour}, {Johnson}, {Poole} \& {Staveley-Smith}}{{Koda}
  et~al.}{2013}]{1312.1022}
{Koda} J.,  {Blake} C.,  {Davis} T.,  {Magoulas} C.,  {Springob} C.~M.,
  {Scrimgeour} M.,  {Johnson} A.,  {Poole} G.~B.,    {Staveley-Smith} L.,
  2013, ArXiv e-prints

\bibitem[\protect\citeauthoryear{{Kopp}, {Uhlemann} \& {Haugg}}{{Kopp}
  et~al.}{2014}]{1312.3638}
{Kopp} M.,  {Uhlemann} C.,    {Haugg} T.,  2014, \jcap, 3, 18

\bibitem[\protect\citeauthoryear{{Lu}, {Ananda} \& {Clarkson}}{{Lu}
  et~al.}{2008}]{0709.1619}
{Lu} T.~H.-C.,  {Ananda} K.,    {Clarkson} C.,  2008, \prd, 77, 043523

\bibitem[\protect\citeauthoryear{{Matarrese} \& {Terranova}}{{Matarrese} \&
  {Terranova}}{1996}]{mataterra}
{Matarrese} S.,  {Terranova} D.,  1996, \mnras, 283, 400

\bibitem[\protect\citeauthoryear{Milillo}{Milillo}{2010}]{thesis}
Milillo I.,  2010, PhD thesis, University of Portsmouth

\bibitem[\protect\citeauthoryear{{Milillo}, {Bertacca}, {Bruni} \&
  {Maselli}}{{Milillo} et~al.}{2015}]{postf}
{Milillo} I.,  {Bertacca} D.,  {Bruni} M.,    {Maselli} A.,  2015, preprint
  (arXiv:1502.02985)

\bibitem[\protect\citeauthoryear{Poisson \& Will}{Poisson \&
  Will}{2014}]{poisandwillbook}
Poisson E.,  Will C.~M.,  2014, {Gravity: Newtonian, Post-Newtonian,
  Relativistic}

\bibitem[\protect\citeauthoryear{{Pueblas} \& {Scoccimarro}}{{Pueblas} \&
  {Scoccimarro}}{2009}]{pueb}
{Pueblas} S.,  {Scoccimarro} R.,  2009, \prd, 80, 043504

\bibitem[\protect\citeauthoryear{{Schaap} \& {van de Weygaert}}{{Schaap} \&
  {van de Weygaert}}{2000}]{dtfe1}
{Schaap} W.~E.,  {van de Weygaert} R.,  2000, \aap, 363, L29

\bibitem[\protect\citeauthoryear{{Shibata} \& {Asada}}{{Shibata} \&
  {Asada}}{1995}]{asada}
{Shibata} M.,  {Asada} H.,  1995, Progress of Theoretical Physics, 94, 11

\bibitem[\protect\citeauthoryear{{Springel}}{{Springel}}{2005}]{gadget2}
{Springel} V.,  2005, \mnras, 364, 1105

\bibitem[\protect\citeauthoryear{{Takada} \& {Futamase}}{{Takada} \&
  {Futamase}}{1997}]{takfut97}
{Takada} M.,  {Futamase} T.,  1997, ArXiv Astrophysics e-prints

\bibitem[\protect\citeauthoryear{{Thomas}, {Bruni}, {Koyama}, {Li} \&
  {Zhao}}{{Thomas} et~al.}{2015}]{dan_fr}
{Thomas} D.~B.,  {Bruni} M.,  {Koyama} K.,  {Li} B.,    {Zhao} G.-B.,  2015,
  ArXiv e-prints

\bibitem[\protect\citeauthoryear{{Thomas}, {Bruni} \& {Wands}}{{Thomas}
  et~al.}{2014}]{1403.4947}
{Thomas} D.~B.,  {Bruni} M.,    {Wands} D.,  2014, preprint (arXiv:1403.4947)

\bibitem[\protect\citeauthoryear{{Tomita}}{{Tomita}}{1991}]{tomitaflat}
{Tomita} K.,  1991, Progress of Theoretical Physics, 85, 1041

\bibitem[\protect\citeauthoryear{{van de Weygaert} \& {Schaap}}{{van de
  Weygaert} \& {Schaap}}{2009}]{dtfe2}
{van de Weygaert} R.,  {Schaap} W.,  2009, in {Mart{\'{\i}}nez} V.~J.,  {Saar}
  E.,  {Mart{\'{\i}}nez-Gonz{\'a}lez} E.,   {Pons-Border{\'{\i}}a} M.-J.,  eds,
  Data Analysis in Cosmology Vol.~665 of Lecture Notes in Physics, Berlin
  Springer Verlag, {The Cosmic Web: Geometric Analysis}.
pp 291--413

\bibitem[\protect\citeauthoryear{{Weinberg}}{{Weinberg}}{1972}]{weinbook}
{Weinberg} S.,  1972, {Gravitation and Cosmology: Principles and Applications
  of the General Theory of Relativity}

\bibitem[\protect\citeauthoryear{{Zheng}, {Zhang}, {Jing}, {Lin} \&
  {Pan}}{{Zheng} et~al.}{2013}]{1308.0886}
{Zheng} Y.,  {Zhang} P.,  {Jing} Y.,  {Lin} W.,    {Pan} J.,  2013, \prd, 88,
  103510

\end{thebibliography}

\end{document}